\documentclass[twocolumn]{aastex63}

\usepackage{float}
\makeatletter\let\newfloat\newfloat@ltx\makeatother
\usepackage{algorithm}
\usepackage{amsmath}
\usepackage{algpseudocode}
\usepackage{booktabs} 
\newcommand{\code}[1]{\texttt{#1}}
\usepackage[]{xcolor}

\received{July 18, 2023}
\accepted{August 20, 2024}

\shorttitle{Pythia}
\shortauthors{Sravan et al.}

\begin{document}

\title{Machine-directed gravitational-wave counterpart discovery}

\author{Niharika Sravan}
\affiliation{Department of Physics, Drexel University, Philadelphia, PA 19104, USA}
\affiliation{Cahill Center for Astrophysics, California Institute of Technology, Pasadena CA 91125, USA}

\author{Matthew J. Graham}
\affiliation{Cahill Center for Astrophysics, California Institute of Technology, Pasadena CA 91125, USA}

\author[0000-0002-8262-2924]{Michael W. Coughlin}
\affiliation{School of Physics and Astronomy, University of Minnesota, Minneapolis, Minnesota 55455, USA}

\author{Tomas Ahumada}
\affiliation{Cahill Center for Astrophysics, California Institute of Technology, Pasadena CA 91125, USA}

\author{Shreya Anand}
\affiliation{Cahill Center for Astrophysics, California Institute of Technology, Pasadena CA 91125, USA}



\begin{abstract}
Joint observations in electromagnetic and gravitational waves shed light on the physics of objects and surrounding environments with extreme gravity that are otherwise unreachable via siloed observations in each messenger. However, such detections remain challenging due to the rapid and faint nature of counterparts. Protocols for discovery and inference still rely on human experts manually inspecting survey alert streams and intuiting optimal usage of limited follow-up resources. Strategizing an optimal follow-up program requires adaptive sequential decision-making given evolving light curve data that (i) maximizes a global objective despite incomplete information and (ii) is robust to stochasticity introduced by detectors/observing conditions. Reinforcement learning (RL)  approaches allow agents to implicitly learn the physics/detector dynamics and the behavior policy that maximize a designated objective through experience.
To demonstrate the utility of such an approach for the kilonova follow-up problem, we train a toy RL agent for the goal of maximizing follow-up photometry for the true kilonova among several contaminant transient light curves. In a simulated environment where the agent learns online, it achieves 3$\times$ higher accuracy compared to a random strategy. However, it is surpassed by human agents by up to a factor of 2. This is likely because our hypothesis function ($Q$ that is linear in state-action features) is an insufficient representation of the optimal behavior policy. More complex agents could perform at par or surpass human experts. Agents like these could pave the way for machine-directed software infrastructure to efficiently respond to next generation detectors, for conducting science inference and optimally planning expensive follow-up observations, scalably and with demonstrable performance guarantees. 
\end{abstract}

\keywords{Gravitational wave astronomy (1043); Time domain astronomy (2109); Surveys (1671); Gravitational wave sources (677); Transient detection (1957)}


\section{Introduction} \label{sec:intro}
The inspiral and subsequent merger of compact objects emits gravitational waves (GWs) that can be detected by ground-based interferometers like LIGO \citep{LIGO_detector}, Virgo \citep{Virgo_detector}, and KAGRA \citep{KAGRA_detector} (also the International Gravitational-wave Network; IGWN). If at least one of the binary components is a neutron star (NS), these phenomena may have analogs in electromagnetic (EM) waves with signatures spanning the EM spectrum, from gamma rays to radio waves. Kilonovae (KNe) are ultraviolet, optical, near-infrared transients that arise due to rapid neutron capture ({\it r}-process) nucleosynthesis in the merger ejecta \citep{1974ApJ...192L.145L, 1982ApL....22..143S, 1998ApJ...507L..59L}. They are robust counterparts to most binary neutron star (BNS) and many neutron-star black hole (NSBH) mergers \citep{2010MNRAS.406.2650M}. Their light curves and spectra provide the opportunity to probe the ejecta and processes driving nucleosynthesis in it \citep{2017Natur.551...80K, 2017Sci...358.1570D, 2017Sci...358.1559K}, and hypermassive/supramassive NS formation scenarios to place independent constraints on the NS equation of state (EoS) \citep{2017ApJ...850L..34B, 2018ApJ...852L..29R}. However, they are quite elusive because they are faint (M$\lesssim$ -16 mag in optical) and short lived ($\lesssim$1 week) \citep{2019MNRAS.489.5037B, 2019LRR....23....1M}.

AT2017gfo was the first, and so far only, kilonova associated with a GW signal \citep[GW170817,][]{2017PhRvL.119p1101A} discovered \citep{2017Sci...358.1556C, 2017Natur.551...75S}. This milestone was achieved due to data in all messengers being broadcast worldwide in real-time and cyberinfrastructure/algorithms for unifying and characterizing them. On the EM side, this includes General Coordinates Network (GCN) notices of gamma ray bursts (GRBs) associated with jets launched by accretion onto the merger remnant \citep{1992ApJ...395L..83N}, new transients discovered by ground-based surveys \citep{2017PASP..129j4502K, 2018PASP..130f4505T, 2014htu..conf...27B, 2017NatAs...1E..71B} broadcast via Astronote/Transient Name Server and data brokers \citep{2021AJ....161..242F, 2019A&A...631A.147N, 2014SPIE.9149E..08S, 2021MNRAS.501.3272M}, and follow-up observations co-ordinated by marshals/Target and Observation Managers \citep{2018SPIE10707E..11S, 2023ApJS..267...31C} and performed by networks of telescopes \citep{2020MNRAS.497.5518A, 2020ApJ...905..145K}. While each of these solutions represents important components of the required cyberinfrastructure for identifying KNe, they are not sufficient. 

When the IGWN detects a signal from a compact binary coalescence, they issue a GCN notice to the astronomy community, along with some GW-derived inference, like source sky localization and distance\footnote{\url{https://emfollow.docs.ligo.org/userguide/}}. Depending on the orientation of the system, satellites could detect a gamma-ray burst (GRB) powered by the merger and help reduce the search area by crossmatching credible regions. In fact, several GRB afterglow target-of-opportunity (ToO) searches \citep{2015ApJ...806...52S,2022ApJ...932...40A} have paved the way for GW ToO campaigns \citep{2020ApJ...905..145K}. Transients that are temporally and spatially coincident with the GW event are assembled. The vast majority of contaminants in this list are supernovae, cataclysmic variables (CVs), and unassociated GRB afterglows. Especially at early phases, these can be difficult to distinguish from KNe. Human experts manually examine the candidate light curves together with additional information, like host galaxies/environment, preliminary fits to light curve models (e.g. kilonovae and common imposters, such as shock cooling emission), and perform additional follow-up to obtain better characterization, and hopefully identify the real event. This step can occur at late hours via oral/written electronic communication and is the one of the main sources of inefficiency in GW counterpart searches.

While key aspects of this protocol have been successful at identifying novel and elusive fast transients like Fast Blue Optical Transients \citep{2022ApJ...932..116H, 2023ApJ...949..120H, 2020ApJ...895...49H} and GRB afterglows \citep{2021ApJ...918...63A}, it has struggled at identifying KNe. This is because these discoveries were only possible when the light curve was sufficiently resolved to allow early estimates for e.g. decline rate. For the vast majority of KNe, there are too few photometric points to reveal any signature characteristics. In fact, post-mortem archival searches have not revealed new KNe \citep{2020ApJ...904..155A}. This indicates that additional real-time follow-up is crucial for identification. At the same time, candidates requiring follow-up far exceed available resources. In practice, especially for KNe at or close to detection limits, human-directed follow-up strategies can be no better than guesswork. 

The situation appears daunting as we begin IGWN's current observing run (O4). Owing to the increased sensitivity from its previous run (O3), the network is projected to discover $\sim 10\times$ more BNS/NSBH mergers than O3 \citep{2018LRR....21....3A, 2022ApJ...924...54P, 2023ApJ...958..158K}. The associated sky localizations are expected to improve, but not by much \citep[e.g. Table 4 in][]{2023ApJ...958..158K}, leaving the effective sky footprint needing to be searched nearly the same. Moreover, only a fraction of BNS mergers are likely to be accompanied by a detectable GRB as with GW170817 \citep{2018Galax...6..130M}, which was key in the rapid identification of AT2017gfo. The search for KNe using current human-centered protocols is ill-suited to handle the increased volume and will continue resulting in inefficiencies and lost opportunities of both human and scientific resources.

Optimal resource allocation has received a lot of interest given the deluge of survey data and limited follow-up resources. To first order, this can be performed using population analyses \citep[e.g.,][]{2018ApJ...867..135C, 2019ApJ...880L..22W, 2020ApJ...889...36C} to identify strategies that optimize a given metric for the entire population instead of being customized to each event. More targeted approaches, e.g. event selection for spectroscopic follow-up, have recently been developed using sophisticated machine learning approaches \citep{2019MNRAS.483....2I, 2020AJ....159...16A, 2020arXiv201005941K, 2022MLS&T...3a5023W}. Reinforcement Learning (RL) was recently used to optimize spectroscopic follow-up of galaxies to maximize Shannon Information on cosmological parameters \citep{2021arXiv210609761C}. In contrast, our work focuses on real-time resource allocation. The use of RL to plan the follow-up of transients is necessary because information gained from follow-up may only be apparent after all the data arrives (discussed in Section \ref{s:back}). RL agents are also robust to stochasticity, which is intrinsic to survey data and the execution of ToOs. These inherent complexities render approaches such as linear programming or rule-based systems inadequate.

In this paper, we demonstrate using a toy RL agent that the KN follow-up problem can be addressed using artificial intelligence (AI). Given $N$ transient light curves from Zwicky Transient Facility (ZTF), one of which is a KN and the rest are contaminants, the agent must maximize an additional follow-up photometry (300s exposure in ZTF $g$, $r$ or $i$) allocated to the true KN each night for 6 nights. Despite its simplicity, our agent learns a good behavior policy and comes close to human-level performance. Our agent belongs to the class of ORACLEs\footnote{Object Recommender for Augmentation and Coordinating Liaison Engine} \citep{2020ApJ...893..127S, 2021arXiv211205897S}.

This paper is organized as follows. In Section \ref{s:back} we introduce elementary concepts in RL. We detail our training data and algorithm in Section \ref{s:pythia}. We present our results in Section \ref{s:results} where we also analyze the strategies of both the AI and human agents. We conclude in Section \ref{s:concl}.
Our code is available at \url{https://github.com/niharika-sravan/Pythia} \citep{niharika_sravan_2024_10995342}.

\section{Autonomous Control} \label{s:back}
To solve sequential decision making under uncertainty problems one needs to consider caveats that make it very different from vanilla machine learning. First, there is no longer a static distribution of data (assumed identically and independently distribute or i.i.d) represented by features $X$, and in supervised learning, target $Y$, where the goal is to the learn a generalized function $f(X)$, that in supervised learning maps to $Y$. The key issue in adaptive control problems is that the distribution of observed data evolves as a result of the AI agent interacting with it and is no longer i.i.d. This is because the actions chosen by the agent determine the next set of observations, making them highly correlated. Using supervised learning in these situations can lead to compounding errors as the data distribution starts deviating from an initial dataset \citep{pmlr-v15-ross11a}. Closely related is the idea of exploration. In vanilla machine learning, one assumes that the training data is all there is and the prediction task will be on a dataset that follows a similar distribution. There is no notion of acquiring more data to improve on the optimization problem. An exception is active learning, where labels from an oracle (e.g. a human expert) are acquired to improve the algorithm's performance. However, even in this case, the training and target data distribution are assumed to remain i.i.d. Another key difference is the notion of delayed consequences. Using the KN follow-up example, an agent may take a decision to obtain follow-up today but it will not be clear until the entire data comes as to how important the resulting data (and therefore how good the decision) was for constraining light curve physics. This makes it hard to solve the so-called ``credit assignment problem'', which is to identify which action among a series of actions produced the desirable outcome.

\begin{deluxetable*}{lcc}
\tablecaption{{{RL formulation used in this work}} \label{t:form}}
\tablewidth{0pt}
\tablehead{\colhead{Parameter} & \colhead{Definition} & \colhead{Choice} }
\startdata
State		& World observed by the agent			& $N$ ZTF transient light curves \\
Action	& Choices presented to the agent		& Obtain single photometric follow-up in ZTF $g$, $r$ or $i$ \\
Reward	& Score the agent receives	 		& 1 if assigned to the KN, 0 otherwise \\
Policy	& Rule specifying action to take			& $\epsilon$-greedy \\
VFA 		& Generalization to new state-actions	& Linear in state-action features \\
Q target 	& Approximation of $Q^*$				& TD(0) \\
Algorithm 	& Learning to update $\hat{Q}$			& SARSA \\
Optimizer & Learning rate						& Adam \\ 
\enddata
\end{deluxetable*}

Here we introduce some background in RL to clarify our work (see \citet{sutton2018reinforcement} for a detailed treatise). To solve RL problems one needs to formulate the control task as a Markov Decision Process (MDP). An MDP is defined by a sequence of states, actions, rewards, next state and so on.

The {\it state} is a mathematical representation of the world that the agent observes. For the KN follow-up problem, this is data contained in GCNs, survey data (light curves and image stamps), and any miscellaneous information relevant to the problem. 
The {\it action} is a mathematical representation of how an agent responds. For the KN photometric follow-up problem, this might be choosing among a set of follow-up observations, e.g. \{no action, observe $g$-band, observe $r$-band, observe $g$-band \& $r$-band\}. Actions may be associated with costs and subject to a budget.
The {\it policy} is a function $\pi(a\mid s) = P(a_t = a \mid s_t = s)$, mapping the agent’s observed state to a distribution of actions it takes. 
The {\it reward} is the utility of taking a given action in a given state. This encodes the science objective needing maximization. For the KN follow-up case, this could be the improvement in constraints on model parameters. Such rewards will be delayed, making the use of RL indispensable for such applications.

In an MDP, at timestep $t$, an agent observes the state of the world $s_t$ returned by the environment and takes an action $a_t$ according to its policy. The environment returns to the agent a new state $s_{t+1}$ along with a reward $r_t$ and the process repeats, either forever or until a horizon, $H$, is reached. 

In many RL problems, one requires the Markov property to be satisfied, in that the state representation is a sufficient statistic of the full history. Specifically,
\begin{equation}
p(s_{t+1} \mid s_t , a_t ) = p(s_{t+1} \mid h_t , a_t )
\end{equation} 
where $s_t$ is the state of the world at time $t$ and $h_t$ is the history of states, actions, and rewards up to timestep $t$. The Markov structure is useful because it allows to formulate the decision problem as a dynamic program with sub-problems that can be optimized locally and to leverage bootstrapped estimates from previous iterations. This relies on a modified form of the Bellman equation for control:
\begin{equation} \label{e:sa}
Q(s_t, a_t) = R(s_t, a_t) + \gamma \sum_{s_{t+1} \in S} P(s_{t+1} \mid s_t, a_t) V(s_t)
\end{equation}
where $Q$ is the state-action value of policy $\pi$, $R(s_t, a_t)$ is the immediate reward from taking action $a_t$ in state $s_t$, $s_{t+1}$ is the next state determined by $P(s_{t+1} \mid s_t, a_t)$, which is the probabilistic transition/dynamics model, and $V$ is the value function, which is the expected discounted sum of future rewards for being in state $s_t$, given by,
\begin{equation}
V (s_t) = E[r_t + \gamma r_{t+1} + \gamma^2 r_{t+2} + \ldots + \gamma^{H-1}r_{t+H-1}\mid s_t = s]
\end{equation}
$H$ is the horizon, which could be infinite. $\gamma$ is the discount factor, $\gamma = 0$ makes the agent myopic, i.e. it only cares about immediate rewards, $\gamma = 1$ means the agent values current and future rewards equally. $\gamma$ controls how fast reward information is propagated to earlier timesteps in value/policy updates. 
The goal of most RL problems is to either learn $Q$ or $\pi$. 

For finite states, the idea is to estimate $Q$ as an iterative update using the Bellman identity. Due to the contraction property of the Bellman operator, such policy iteration algorithms converge to the optimal action-value function, $Q^*$, as $i \to \infty$, where $i$ is the iteration count. 
In most RL problems, it is computationally infeasible to compute the expectation over next states (denoted by the sum in Equation \ref{e:sa}). Therefore, it is common to sample the expectation using Monte Carlo or Temporal Difference (TD) methods \citep{10.1145/203330.203343}. 
Notably, TD methods both sample and bootstrap by approximating the expectation as $Q(s_{t+1}, a_{t+1})$ in SARSA \citep{Rummery1994OnlineQU} or $\max_a Q(s_{t+1}, a)$ in Q-learning \citep{watkins1989learning}, where Q is the bootstrapped using the running estimate. TD estimates are biased (as it bootstraps) and have high variance. However, TD methods are more versatile than Monte Carlo as they can handle infinite horizon MDPs. In this work, we use TD methods.

In most realistic situations, where the state space is continuous, such an iterative update is not computationally tractable since the cardinality of $Q(s, a)$ is infinite. In practice, it is therefore common to use a function approximator to estimate $Q(s, a; \theta)$, where $\theta$ are parameters describing $Q$. In RL, traditionally this is a linear value function approximator \citep[VFA,][]{10.1145/203330.203343}, where $Q$ is parameterized by linear weights ($w$) as a function of state-action features, i.e. $Q(s, a; w)=x(s,a)^T w$. More recently, non-linear function approximators (like neural networks) have been used \citep{Mnih2013PlayingAW} with much success. Updates to $Q$ are then achieved using stochastic gradient descent along the mean squared error between the Q approximation (e.g. TD estimate) and the running estimate.  
In this work, we adopt linear VFA in SARSA despite its simplicity because the algorithm is guaranteed to converge to within a constant factor of the optimal behavior policy under minor assumptions \citep[][discussed in detail in Section \ref{s:algo}]{580874}. This is not the case for Q-learning or non-linear approximators. This attribute makes it suitable for our pilot investigation where we are interested in exploring the suitability of RL approaches to solve the KN follow-up problem and providing useful benchmarks on the same. 

\begin{algorithm*}
\caption{SARSA and TD(0) target}\label{alg:rl}
\begin{algorithmic}
\State Initialize $w$ to small random weights 
\State Set $\epsilon_0 = 1$
\For{k = 1, M} \Comment For each episode
	\State $\epsilon \gets \epsilon_0/k^{n}$
	\State Initialize $s_1$
	\For{t = 1, horizon}
		\State With probability $\epsilon$ select random action $a_t$
		\State otherwise select $a_t = \max_a \hat{Q}(s_t,a_t; \hat{w})$
		\State Execute action and observe reward $r_t$ and next state $s_{t+1}$ from environment
		\State With probability $\epsilon$ select random action $a_{t+1}$
		\State otherwise select $a_{t+1} = \max_a \hat{Q}(s_{t+1},a_{t+1}; \hat{w})$
		\State Set $\Delta \hat{w} \gets [r_t+ \gamma \hat{Q}(s_{t+1}, a_{t+1}; \hat{w}) - \hat{Q}(s_{t}, a_{t}; \hat{w})] \nabla_w \hat{Q}(s_{t}, a_{t}; \hat{w})$ 
		\State \Comment Loss is MSE between TD(0) target (substitute for $Q^*$) and current $Q$ 
	\State Update $\hat{w} \gets \hat{w} +  \alpha \Delta \hat{w}$ \Comment $\alpha$ is using Adam
	\EndFor
\EndFor
\end{algorithmic}
\end{algorithm*}

We note that, in the setup of an MDP and solutions via RL, traditional supervised learning approaches may be useful in several places. A choice of a sufficient statistic for the full history of a state might be a Long short-term memory (LSTM) encoding of the time-series, the dynamics model may be conditional probabilistic forecasts of the time series, the reward might be constraints on the physics model using an emulator.

\section{\texttt{Pythia}: a KN photometric follow-up agent} \label{s:pythia}
The KN follow-up problem requires a deterministic reward, infinite state (light curves are continuous functions), and finite horizon (KNe are transients and have finite lifetimes) MDP. Here we focus on deterministic and discrete action domains. We note that, in practice, follow-up actions can be stochastic, e.g. if the strategized observation is not executed (e.g. due to weather) or executed erroneously. Ideally the action space is also continuous and subject to an overall budget. Observing time is often awarded as cumulative hours over an observing season with each observation associated with a different cost in exposure time. In the following, we describe our RL setup (summarized in Table \ref{t:form}).

\subsection{Problem Statement} \label{s:prob}
Our agent is presented $N$ transient light curves from ZTF, one of which is the KN and the rest are contaminants, chosen randomly from a list of supernovae and unassociated GRB afterglows. The agent observes the candidates on day 1. On days 2 through 7 it assigns one additional photometry point with ZTF in $g$, $r$, or $i$ using a deep 300s exposure to one of the events. The agent gets a reward 1 if the follow-up is assigned to the KN, and 0 otherwise. The objective is to learn a policy to maximize the number of follow-up assigned to the true KN. The maximum achievable score is 6. A random agent will achieve an expected score of $6/N$. 

We have chosen a non-model specific reward that seeks to maximize sampling, with the expectation that more data leads to better constraints. KNe are quite novel and there are still a lot of open questions surrounding ejecta composition and the treatment of opacities and its effect on the KN light curve \citep{2019LRR....23....1M}. Targeting constraining models could then result in poor generalizability or transfer to real-world domains of learned behavior policies. However, given reliable and representative models, constraining physics parameters could be a good choice.

\subsection{Training Data}
We generate training light curves as follows. We place transients randomly in the 3-D volume of the Universe observable by ZTF.
ZTF is a public-private partnership survey of the Northern sky conducted in $g$, $r$, and $i$ filters using a wide field (47 deg$^2$) imager on the Palomar 48-inch telescope \citep{2019PASP..131a8003M, 2020PASP..132c8001D, 2019PASP..131g8001G, 2019PASP..131a8002B}.
For Type Ia supernovae we use SALT2 \citep{2007A&A...466...11G} models to estimate photometry in ZTF filters. We apply Milky way extinction \citep{1989ApJ...345..245C,1998ApJ...500..525S} given the supernova sky location, redshift, and K-corrections.
For core-collapse supernovae, we use 2-D Gaussian Process fits to ZTF light curves of spectroscopically confirmed supernovae of types II, IIn, IIb, Ib, Ic, Ic-BL, and superluminous supernovae types I and II. While we apply redshift corrections (scaling and stretch), due to the empirical nature of the fit, we are unable to apply extinction corrections. We estimate K-correction by estimating the Gaussian Process fit in the wavelength dimension at the redshifted mean wavelength of the ZTF filter at the sampled redshift. 
For GRB afterglows, we use the semi-analytic models of \cite{2020ApJ...896..166R}. 
Finally, we use the models from \citet{2019MNRAS.489.5037B} to simulate KNe from BNS and NSBH mergers. For KNe we use the interpolation scheme in \citet{2023NatCo..14.8352P}.
For these, once again, while redshift scaling is accounted for, we do not account for extinction or K-correction. 

We sample the above light curves according to ZTF Phase-II public/private $i$-band cadence and ToO observations performed the first day or first two days after GW trigger according to \citet{2021NatAs...5...46A} (see \citet{2023ApJ...958..158K} for details).
For Type Ia supernovae, we also add noise estimated using SALT2 fits to classified ZTF Type Ia supernovae. 
Simulations last up to 7 days after trigger. For each type mentioned above we generate 10,000 light curves in a log redshift space for supernovae and uniform space in luminosity distance for the rest. The former ensures more uniform training data as a function of peak supernova magnitude.

\begin{figure} \centering 
\includegraphics[width=\linewidth]{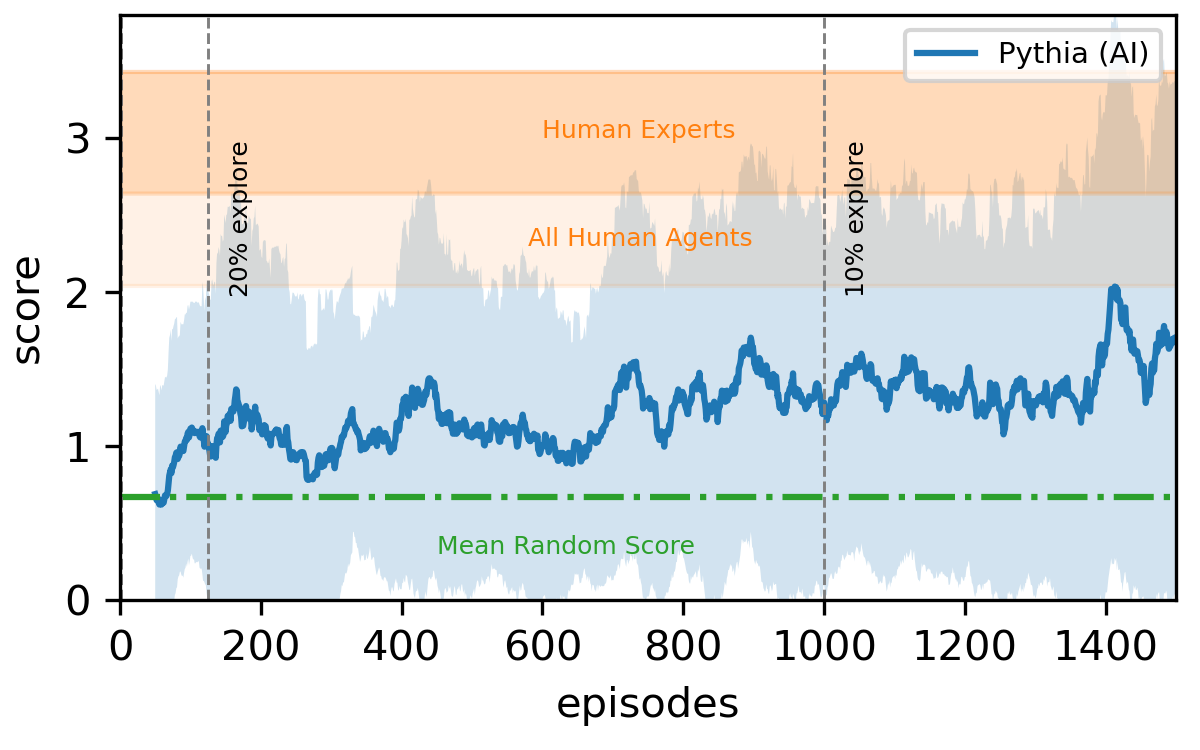}
\caption{Learning curve of our KN follow-up agent \texttt{Pythia} (blue). Our agent achieves $> 3.5\times$ higher score than a random agent during training. The score variance (shaded blue region) is computed using a moving average with a length of 50 episodes. The mean random score is 6/9. Human references are shown as orange shaded regions.}
\label{f:pythia}
\end{figure}

\begin{figure*} \centering 
\includegraphics[width=\linewidth]{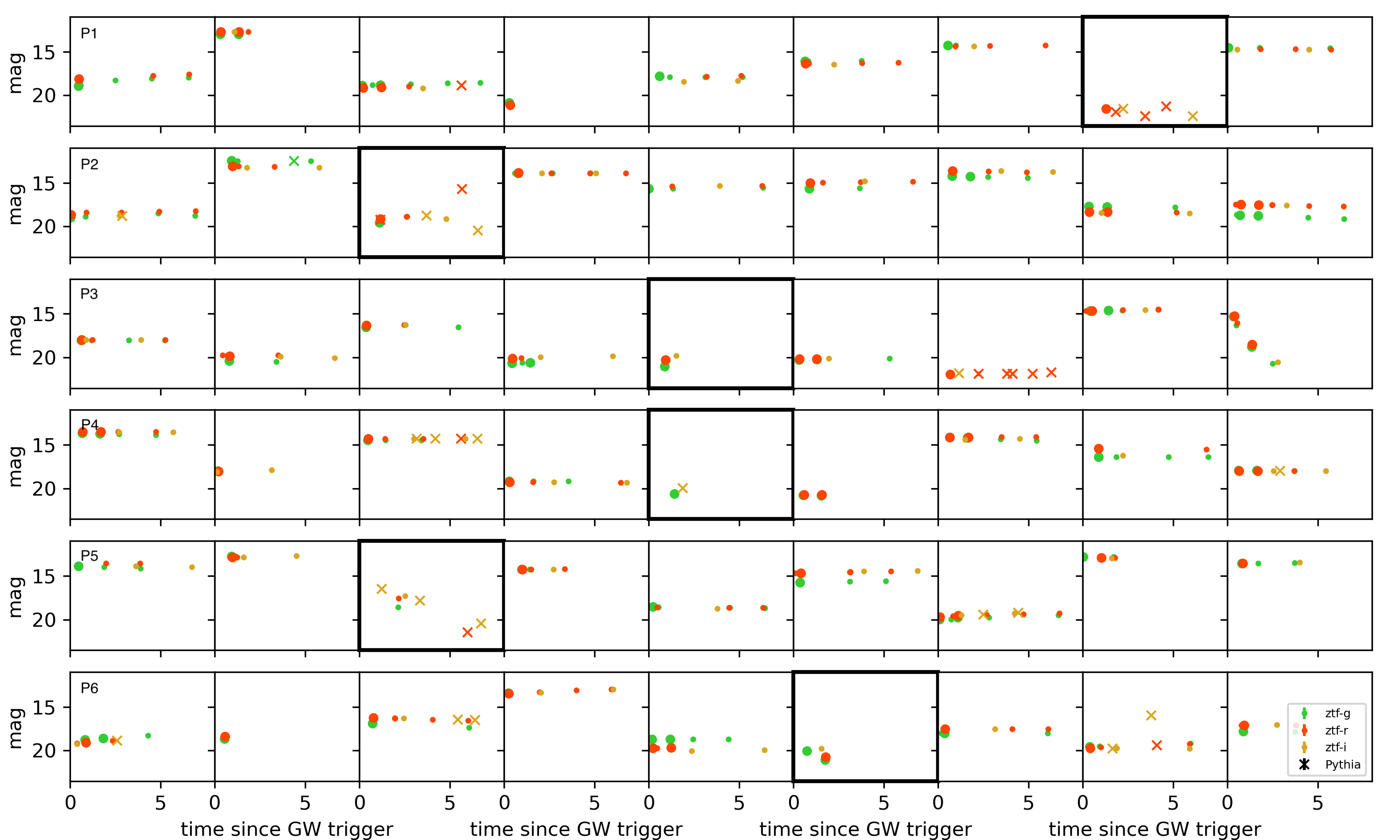}
\caption{Random sample of test scenarios with 9 events, one of which is the true KN (emphasized in black borders). Forced photometry light curves from ZTF are shown in circles and follow-up photometry allocated by \texttt{Pythia} are shown as crosses. The goal is to maximize follow-up for the true KN. The agent is not informed which the true event is.}
\label{f:sample-pythia}
\end{figure*}

\begin{figure*} \centering 
\includegraphics[width=\linewidth]{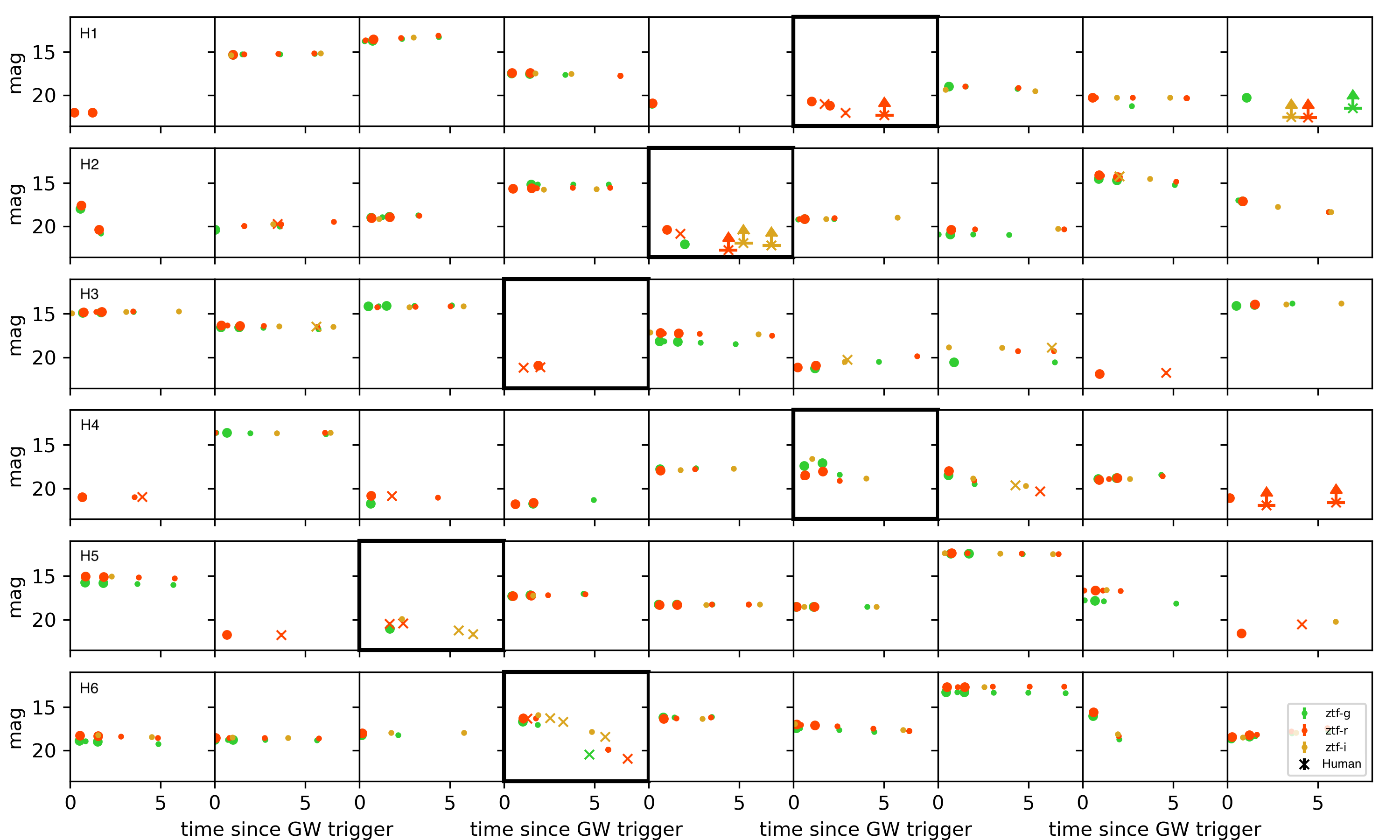}
\caption{Same as Figure \ref{f:sample-pythia} but showing follow-up photometry allocated by the best performing human agent (Expert 4).}
\label{f:sample-human}
\end{figure*}

\subsection{RL Algorithm} \label{s:algo}
We learn the optimal behavior policy using SARSA. SARSA is a simple on-policy learning algorithm, i.e. samples from the policy are used to update $Q$ (see Algorithm \ref{alg:rl}). We parametrize $Q =x(s,a)^T w$, where $x(s,a)$ is the state-action feature representation and $w$ is a set of linear weights to be learned. We choose a TD(0) target, i.e. $r_t+ \gamma \hat{Q}(s_{t+1}, a_{t+1}; \hat{w})$, where $\hat{Q}$ is the running estimate. The policy is $\epsilon$-greedy: with probability $\epsilon$ the agent chooses a random action and otherwise chooses $\max_a \hat{Q}(s_t,a_t; \hat{w})$, the action with the maximum $\hat{Q}$-value. If the learning rate $\alpha_t$ satisfies the Robbins-Munro sequence and $\epsilon \to 0$ as $i \to \infty$ the agent is Greedy in the Limit of Infinite Exploration (GLIE) and is a sufficient condition for convergence in SARSA\footnote{Two conditions to be GLIE: (1) we visit every state-action pair infinite times and (2) behavior policy converges to greedy policy}. We parameterize $\epsilon = 1/k^n$, where $n$ is a hyperparameter that controls how fast the agent becomes greedy. This parameterization allows the agent to get in a lot of exploration early during training and gradually forces it become more greedy. To ensure sample efficiency and propagate TD updates faster, we reuse episodes $m$ times, where we set $m=5$.
Our agent learns online (i.e. it collects new experiences as opposed to episodes collected beforehand) in a simulated environment\footnote{It is possible to train the agent online with real GW trigger scenarios but RL algorithms are notoriously sample inefficient and can take a lot of exploration to arrive at the optimal $Q$/policy.}. 

We estimate the distribution of forecast photometry given an action using 2-D Gaussian Process regression on the observed light curve. To compute {\it state-action} features, $x(s,a)$, we convert the observed ZTF light curves and forecast photometry given action $a$ to a tiled $3 \times \sqrt N \times \sqrt N$ image that we pass to a convolutional autoencoder (CAE). This choice is important because the agent's decisions need to be invariant to the order of the events. CAEs are appropriate because the convolution operation makes them translation invariant. We choose $x(s,a)$ to be the penultimate layer in the Xception network \citep{2016arXiv161002357C}. This choice precludes us from having to train a custom encoder using state-action pairs derived from a preset policy, which could potentially bias $Q$ estimates. Note that this two-step learning issue (one for state-action features and the other for $Q$) can be avoided using deep Q-networks \citep{Mnih2013PlayingAW}. The observed photometry ({\it next state}) given an {\it action} is simulated using 2-D Gaussian Process regression using either the full ZTF or theoretical model light curves for SNe and GRB afterglows/KNe, respectively. Then, uncertainties and limits are applied in a similar way to the generation of the training dataset.

Our choice of linear VFA was motivated by the fact that the triad of non-linear function approximators, off-policy learning (like Q-learning), and bootstrapping (as in Temporal-difference learning) can cause the $Q$ function to diverge \citep{580874}. Though recent works have made significant progress to address this issue (e.g. prioritized replay \citep{2015arXiv151105952S}, asynchronous methods \citep{2016arXiv160201783M}, dueling DQN targets \citep{2016arXiv161101224W}) we adopt linear VFA and SARSA for our toy implementation to assess the feasibility of RL approaches to solve the KN follow-up problem and disambiguate limitations from convergence issues. An issue may be that the choice of linear VFA as hypothesis function class may not be a sufficiently rich representation of the true $Q$ function. This does indeed appear to be the case as we discuss later.

We refer to our AI agent as \texttt{Pythia}\footnote{Named after the Oracle of Delphi}.

\begin{table}
\centering
\caption{AI and human agent performance} 
\begin{small}
\begin{sc}
\begin{tabular}{|l|c|c|}
\toprule
Agent 		& Score	& Fraction KNe  \\
\midrule
\texttt{Pythia}	& 1.84	& 0.81 \\
Non-expert 1	& 2.04	& 0.54 \\
Non-expert 2	& 3.15 	& 0.86 \\
Expert 1		& 2.64	& 0.76 \\
Expert 2		& 2.74	& 0.78 \\
Expert 3		& 2.94	& 0.72 \\
Expert 4		& 3.43	& 0.90 \\
\bottomrule
\end{tabular} \label{t:perf}
\end{sc}
\end{small}
\end{table}

\section{Results} \label{s:results}

\begin{figure} \centering 
\includegraphics[width=0.95\linewidth]{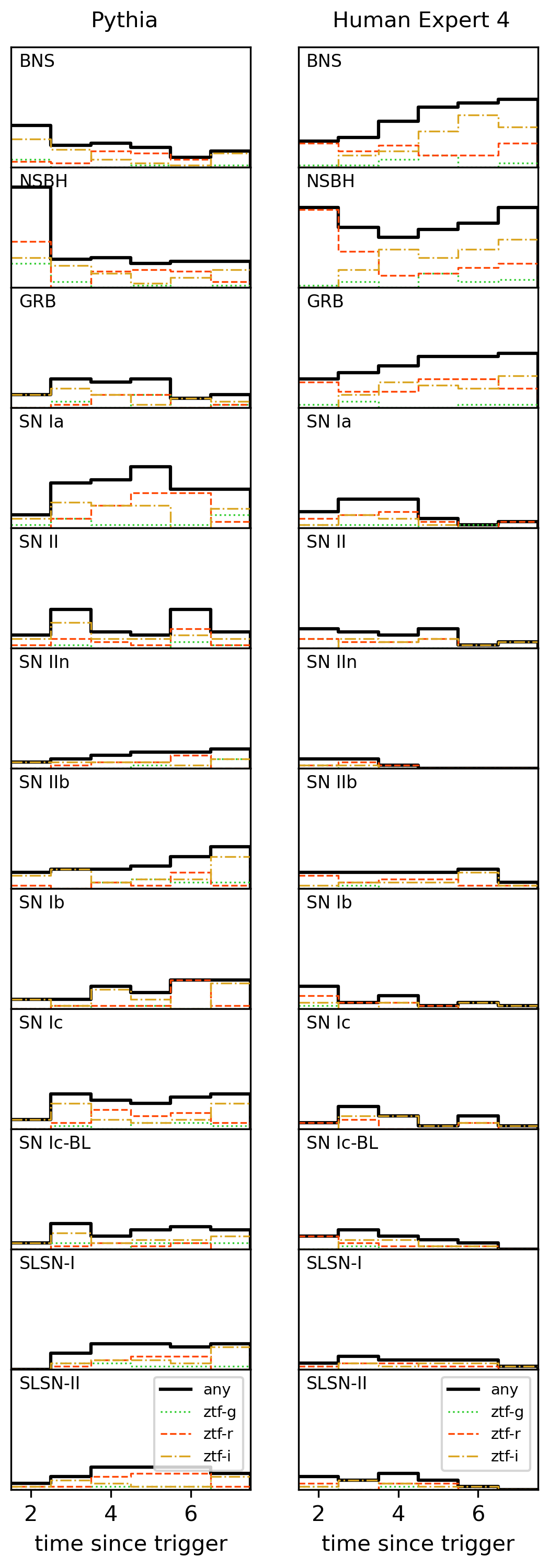}
\caption{Distribution of follow-up allocations at each timestep by transient type according to \texttt{Pythia} (left) and Expert 4 (right). Both AI and human agents preferred redder filters for follow-up.}
\label{f:dist}
\end{figure}

We train \texttt{Pythia} for $N = 9$ events. We perform a grid search for the hyperparameters $\gamma=$ 1.0, 0.5, 0.1, $n=$ 1, 2, 3, and $\alpha=$ 0.1, 0.01, 0.001, 0.0001. Figure \ref{f:pythia} shows the learning curve of our trained agent, smoothed using a moving average with length = 50 episodes, and $\gamma = 0.5$, $n=3$, and $\alpha = 0.01$, our best hyperparameters. \texttt{Pythia} achieves $> 3.5\times$ higher score than a random agent during training. Figure \ref{f:sample-pythia} shows a sample of test episodes with photometry allocated by \texttt{Pythia}. It is 3$\times$ better than random during testing.

\subsection{Human benchmark}
Due to the absence of ground truth in RL problems (the true optimal $Q$ in our case) it is important to compare against strong benchmarks to assess the quality of learned policies. Here we use benchmarks provided by astronomers as a stronger point of reference than a random policy. 

We solicited the aid of six volunteer astronomers to solve our KN follow-up problem. Similar to \texttt{Pythia} they completed a training and testing phase, the former with as many episodes as they chose and the latter with 100 episodes. During both phases, we recorded their choices and decision times. Finally, we asked them to self-identify as experts and non-experts in KN searches and provide a short description of their strategies (summarized in Appendix \ref{s:strat}). 

We list the comparison of the performance of all agents in Table \ref{t:perf}. Score is the average number of photometry allocated to the true KN. Fraction KNe denotes the fraction of episodes in which the true KN received at least one follow-up. The scores for human agents are shown in orange in Figure \ref{f:pythia}. Figure \ref{f:sample-human} shows a sample of episodes with photometry allocated by Expert 4. All human agents surpassed \texttt{Pythia}. Interestingly, \texttt{Pythia} performed well in the fraction of KNe allocated at least one follow-up observation even though it was not the follow-up objective. 

\subsection{Agent preference/policy interpretation}
As with most machine learning approaches, it is hard to interpret what a model has learned. For RL algorithms it can be hard to understand why an agent chose an action in a given situation. Here we try to discern the agents' strategies and understand situations that resulted in failures for them. 

The first point of difference is the agent's behavior when they perform poorly. In such cases, \texttt{Pythia} allocates the bulk of the budget to a single event (see episode P3 and P4 in Figure \ref{f:sample-pythia}). On the other hand, humans spread their budget (see episode H3 in Figure \ref{f:sample-human}), sampling several events to try to rule them out (the intent was communicated to us in discussions after the test). The AI does not explore in failure cases simply because it is not rewarded for doing so. During training, when it makes the wrong choice, the weights do not get updated by much, so it is biased by the weight initialization/experience thus far. In these cases, not updating the target event might be an effective strategy as in high reward/success cases the agent needs to stick with the same event. This aspect underscores the importance of carefully formulating the reward in the design of artificial agents. 

Figure \ref{f:dist} shows the distribution of \texttt{Pythia}'s and the best performing human agent's follow-up allocations by transient type. \texttt{Pythia} is likely to confuse KNe with SNe Ia while the human agent confuses it with GRB afterglows. Note that when simulating our episodes we randomly sample from each type, so the bias toward SNe Ia is not due to observational significance. The human agent performed better for KNe associated with NSBH than BNS mergers. As time since the GW trigger increased, they were more likely to confuse KNe from BNS mergers with GRB afterglows. All humans expressed in the description of their strategies that the decay rate on individual bands was key in deciding whether a light-curve was associated with a merger.
The fast decaying light curves of GRB afterglows paired with their red colors and energetic X-ray/radio counterparts make them notorious KN impostors during human candidate vetting. The cadence with which targeted ToO searches are conducted is highly conducive to GRB afterglow discovery, as the strategy aims for a daily cadence with at least two filters per night. In our test scenarios, human agents had to distinguish between GRB afterglows and KNe by preferentially selecting objects that appeared to redden in color evolution but not decay extremely quickly like GRB afterglows. However, in reality, human vetters are trained to identify the reddest and fastest-fading objects for follow-up. Since GRB afterglows are intrinsically more energetic, their light curve is more populated and allows to observe their fast decay. Even ZTF’s dedicated automated fast transient discovery pipeline, ZTF Realtime Search and Triggering (ZTFReST), did not find any KNe successfully. Instead, the four sources passing the science validation criteria were in fact likely or confirmed GRB afterglows \citep{2021ApJ...918...63A}. Since GRB afterglows are also rare relative to other transient subtypes, there is a higher risk appetite to trigger larger telescope resources on red, fast-fading candidates, even if they may be GRB afterglows.

The KN light curve evolution is governed by multiple parameters. In the \citet{2019MNRAS.489.5037B} models, the wind and dynamical ejecta power the electromagnetic emission, the inclination angle affects the color evolution and peak brightness, the half-opening angle of the lanthanide-rich component affects mainly the color evolution, and the luminosity distance affects the observed brightness.
As such, both AI and human agents preferred redder filters for follow-up. This makes sense as the KN emission peaks in the near infrared, so the light curve will be longer lived in redder filters. The best performing human agent preferred $r$-band follow-up earlier and $i$-band later. However, the specific follow-up strategies with regard to filter choice were different for different human agents (e.g. a different agent preferred $g$-band follow-up first). Note that while the reward was the same regardless of which filter was used to obtain follow-up, it could affect overall the score because the information revealed using it could help in decision-making in the next timestep. Human agents performed worse for episodes that had a merger with small amounts of wind ejecta (i.e $M_{\rm ej,wind} < 10^{-2} M_\odot$) and for which the light curve showed a consistent blue color (e.g. $r-g < 0$). These blue KNe were not picked up by human experts as most were intuitively looking for red KNe, or for a transient evolving into redder colors (see episode H4 in Figure \ref{f:sample-human}). Note however that some of the simulated KN light curves have combinations $M_{\rm ej,wind}$ and $M_{\rm ej,dyn}$ that are inconsistent with numerical relativity simulations (due to extrapolation below grid values).


\section{Conclusions} \label{s:concl}
In this paper we develop the first AI agent capable of strategizing a sequential transient follow-up program. Given several events with evolving light curves, one of which is a KN and the rest are contaminants, the agent must maximize follow-up for the true candidate, using information learned from actions taken and routine survey observations to improve its strategy in the next round. Using a simple behavior policy (linear VFA) our agent demonstrates that the problem is learnable by machines and comes close to human performance. However, the agent is quite greedy in its decisions (the score does not improve much with time) suggesting that the agent's hypothesis function is an insufficient representation of the optimal behavior policy. More complex agents (e.g. using deep $Q$ networks or policy gradient algorithms) could help bridge the gap with human experts. This could really help streamline KN search efforts. 
In the development of such agents it would also be useful to employ graph neural networks to represent the light curves as this is a more natural choice to ensure learning order invariant policies than the convolutional neural networks used here.

An RL agent could assist in KN searches by devising follow-up that is flexible to cater to a variety of objectives. One low-stakes design would be to reward an agent to acquire follow-up that could cull the list of targets under consideration. RL agents would perform better than classifiers alone in the KN discovery process, not only because they take the future into account when deciding the follow-up plan, but also because such decisions would be robust to data acquisition failures due to weather/instrumentation or manual overrides in executing follow-up recommendations.
In order to assist in realistic searches, our framework would need to accommodate variable number of events (candidates are continuously eliminated due to information from other sources), a comprehensive state (e.g. low latency data from IGWN) and action space (more than one photometric point or spectroscopic follow-up, etc along with consideration of observing costs), and return for physics models (e.g. ejecta mass, inclination angle). In fact, as discussed earlier, for optimizing physics constraints in real-time, RL agents are the only solution. 
Finally, trained agents would need to be deployed into existing frameworks (e.g. SkyPortal \citep{2023ApJS..267...31C} or REFITT \citep{2020ApJ...893..127S}) to iteratively execute observing plans and ingest observations.

\begin{acknowledgments}
We thank our reviewer for their careful reading of the manuscript and insightful comments.
We thank the experts from the GROWTH team who helped provide benchmarks reported here. This research was partly completed on Expanse at the San Diego Supercomputer Center at UC San Diego (Accelerate ACCESS Award\# AST200029). 
N.~Sravan acknowledges support from the National Science Foundation with grant numbers AST-2307374.
M.~J.~Graham acknowledges support from the National Science Foundation with grant numbers AST-2034437 and OAC-211799. 
M.~W.~Coughlin was supported by the National Science Foundation award PHY-2308862 and PHY-2117997.
S.~Anand acknowledges support from the National Science Foundation GROWTH PIRE grant No. 1545949.

\end{acknowledgments}

%

\vspace{5mm}


\software{\code{astropy} \citep{astropy:2013, astropy:2018, astropy:2022}, \code{pandas} \citep{mckinney-proc-scipy-2010, reback2020pandas}, \code{tensorflow} \citep{tensorflow2015-whitepaper}, \code{keras} \citep{chollet2015keras}, \code{scikit-learn} \citep{scikit-learn}}


\appendix
\section{Strategies of human agents} \label{s:strat}

Here we summarize the strategies used by human agents who helped provide benchmarks.

\paragraph{Non-expert 1}
During training, the agent noted that KN photometry in the $r$-filter appeared initially brighter than in $g$, which helped them rule out several events early on. They felt they got enough information by monitoring in $r$ and $g$ filters, though they preferred the $r$ filter. They also observed that KN light curves had a moderate decline rate compared to non-KNe, helping them reject events that were either too rapidly or slowly declining. By the second or third day, they needed to observe a significant decrease in brightness to continue choosing the same event. 

\paragraph{Non-expert 2}
The agent focused on light curves showing gradual brightening in the $r$ and $i$ filters and a fade rate $\sim$0.4 mag/day. During the training phase, they observed that transients fading faster than 1 mag/day were GRB afterglows, leading them to exclude those from further follow-up. They also prioritized events with a rapid increase in brightness in the first $2-3$ days.

\paragraph{Expert 1}
The agent primarily focused on identifying fast-evolving events. Once such an event was identified, they observed it repeatedly to monitor its behavior.

\paragraph{Expert 2}
The agent’s initial choice was random between any red transient. They later focused on events with fade rates that were neither too fast (indicative of GRB afterglows) nor too slow (indicative of normal SNe). They were interested in events with an unusual color evolution. They also preferred late observations of single detection events to rule them out.

\paragraph{Expert 3}
The agent's strategy involved first allocating follow-up to the faintest event that had at least one detection or red colors, in order to rule out slowly evolving ones. They then considered color, apparent magnitude, and decay rate, generally ignoring events brighter than $\sim$15 mag (unless displaying rapid fading). They prioritized observations in the $i$-filter as they expected KNe to last longer there.

\paragraph{Expert 4}
The agent prioritized events without initial data, starting with the $r$ filter, because KNe are expected to be fainter and decay faster in $g$. They relied on the $r$-band as their previous experience indicated that observations in $i$ are shallower than $r$. Their main strategy was to measure the decay rate, focusing on events fading faster than 0.3 mag/day with at least two detections. They differentiated between KNe and GRB afterglows based on decline rate and color, rejecting events fading faster than $\sim$1 mag/day. After day 3, they preferred follow-up in the $i$-filter as KNe are expected to be brighter there during later phases. They identified KNe that were consistently red and fading. When this behavior was not observed, the expert would start ruling out events that (1) became brighter, and (2) were consistently blue.

\bibliography{../../KNe, ../../SNIa, ../../ML, ../../ZTF, ../../software}

\begin{thebibliography}{}
\expandafter\ifx\csname natexlab\endcsname\relax\def\natexlab#1{#1}\fi
\providecommand{\url}[1]{\href{#1}{#1}}
\providecommand{\dodoi}[1]{doi:~\href{http://doi.org/#1}{\nolinkurl{#1}}}
\providecommand{\doeprint}[1]{\href{http://ascl.net/#1}{\nolinkurl{http://ascl.net/#1}}}
\providecommand{\doarXiv}[1]{\href{https://arxiv.org/abs/#1}{\nolinkurl{https://arxiv.org/abs/#1}}}

\bibitem[{Abadi {et~al.}(2015)Abadi, Agarwal, Barham, Brevdo, Chen, Citro,
  Corrado, Davis, Dean, Devin, Ghemawat, Goodfellow, Harp, Irving, Isard, Jia,
  Jozefowicz, Kaiser, Kudlur, Levenberg, Man\'{e}, Monga, Moore, Murray, Olah,
  Schuster, Shlens, Steiner, Sutskever, Talwar, Tucker, Vanhoucke, Vasudevan,
  Vi\'{e}gas, Vinyals, Warden, Wattenberg, Wicke, Yu, \&
  Zheng}]{tensorflow2015-whitepaper}
Abadi, M., Agarwal, A., Barham, P., {et~al.} 2015, {TensorFlow}: Large-Scale
  Machine Learning on Heterogeneous Systems.
\newblock \url{https://www.tensorflow.org/}

\bibitem[{{Abbott} {et~al.}(2017){Abbott}, {Abbott}, {Abbott}, {Acernese},
  {Ackley}, {Adams}, {Adams}, {Addesso}, {Adhikari}, {Adya}, {Affeldt},
  {Afrough}, {Agarwal}, {Agathos}, {Agatsuma}, {Aggarwal}, {Aguiar}, {Aiello},
  {Ain}, {Ajith}, {Allen}, {Allen}, {Allocca}, {Altin}, {Amato}, {Ananyeva},
  {Anderson}, {Anderson}, {Angelova}, {Antier}, {Appert}, {Arai}, {Araya},
  {Areeda}, {Arnaud}, {Arun}, {Ascenzi}, {Ashton}, {Ast}, {Aston}, {Astone},
  {Atallah}, {Aufmuth}, {Aulbert}, {AultONeal}, {Austin}, {Avila-Alvarez},
  {Babak}, {Bacon}, {Bader}, {Bae}, {Bailes}, {Baker}, {Baldaccini},
  {Ballardin}, {Ballmer}, {Banagiri}, {Barayoga}, {Barclay}, {Barish},
  {Barker}, {Barkett}, {Barone}, {Barr}, {Barsotti}, {Barsuglia}, {Barta},
  {Barthelmy}, {Bartlett}, {Bartos}, {Bassiri}, {Basti}, {Batch}, {Bawaj},
  {Bayley}, {Bazzan}, {B{\'e}csy}, {Beer}, {Bejger}, {Belahcene}, {Bell},
  {Berger}, {Bergmann}, {Bernuzzi}, {Bero}, {Berry}, {Bersanetti}, {Bertolini},
  {Betzwieser}, {Bhagwat}, {Bhandare}, {Bilenko}, {Billingsley}, {Billman},
  {Birch}, {Birney}, {Birnholtz}, {Biscans}, {Biscoveanu}, {Bisht}, {Bitossi},
  {Biwer}, {Bizouard}, {Blackburn}, {Blackman}, {Blair}, {Blair}, {Blair},
  {Bloemen}, {Bock}, {Bode}, {Boer}, {Bogaert}, {Bohe}, {Bondu}, {Bonilla},
  {Bonnand}, {Boom}, {Bork}, {Boschi}, {Bose}, {Bossie}, {Bouffanais}, {Bozzi},
  {Bradaschia}, {Brady}, {Branchesi}, {Brau}, {Briant}, {Brillet}, {Brinkmann},
  {Brisson}, {Brockill}, {Broida}, {Brooks}, {Brown}, {Brown}, {Brunett},
  {Buchanan}, {Buikema}, {Bulik}, {Bulten}, {Buonanno}, {Buskulic}, {Buy},
  {Byer}, {Cabero}, {Cadonati}, {Cagnoli}, {Cahillane}, {Calder{\'o}n
  Bustillo}, {Callister}, {Calloni}, {Camp}, {Canepa}, {Canizares}, {Cannon},
  {Cao}, {Cao}, {Capano}, {Capocasa}, {Carbognani}, {Caride}, {Carney},
  {Carullo}, {Casanueva Diaz}, {Casentini}, {Caudill}, {Cavagli{\`a}},
  {Cavalier}, {Cavalieri}, {Cella}, {Cepeda}, {Cerd{\'a}-Dur{\'a}n},
  {Cerretani}, {Cesarini}, {Chamberlin}, {Chan}, {Chao}, {Charlton}, {Chase},
  {Chassande-Mottin}, {Chatterjee}, {Chatziioannou}, {Cheeseboro}, {Chen},
  {Chen}, {Chen}, {Cheng}, {Chia}, {Chincarini}, {Chiummo}, {Chmiel}, {Cho},
  {Cho}, {Chow}, {Christensen}, {Chu}, {Chua}, {Chua}, {Chung}, {Chung},
  {Ciani}, {Ciolfi}, {Cirelli}, {Cirone}, {Clara}, {Clark}, {Clearwater},
  {Cleva}, {Cocchieri}, {Coccia}, {Cohadon}, {Cohen}, {Colla}, {Collette},
  {Cominsky}, {Constancio}, {Conti}, {Cooper}, {Corban}, {Corbitt},
  {Cordero-Carri{\'o}n}, {Corley}, {Cornish}, {Corsi}, {Cortese}, {Costa},
  {Coughlin}, {Coughlin}, {Coulon}, {Countryman}, {Couvares}, {Covas}, {Cowan},
  {Coward}, {Cowart}, {Coyne}, {Coyne}, {Creighton}, {Creighton}, {Cripe},
  {Crowder}, {Cullen}, {Cumming}, {Cunningham}, {Cuoco}, {Dal Canton},
  {D{\'a}lya}, {Danilishin}, {D'Antonio}, {Danzmann}, {Dasgupta}, {Da Silva
  Costa}, {Dattilo}, {Dave}, {Davier}, {Davis}, {Daw}, {Day}, {De}, {DeBra},
  {Degallaix}, {De Laurentis}, {Del{\'e}glise}, {Del Pozzo}, {Demos}, {Denker},
  {Dent}, {De Pietri}, {Dergachev}, {De Rosa}, {DeRosa}, {De Rossi}, {DeSalvo},
  {de Varona}, {Devenson}, {Dhurandhar}, {D{\'\i}az}, {Dietrich}, {Di Fiore},
  {Di Giovanni}, {Di Girolamo}, {Di Lieto}, {Di Pace}, {Di Palma}, {Di Renzo},
  {Doctor}, {Dolique}, {Donovan}, {Dooley}, {Doravari}, {Dorrington},
  {Douglas}, {Dovale {\'A}lvarez}, {Downes}, {Drago}, {Dreissigacker},
  {Driggers}, {Du}, {Ducrot}, {Dudi}, {Dupej}, {Dwyer}, {Edo}, {Edwards},
  {Effler}, {Eggenstein}, {Ehrens}, {Eichholz}, {Eikenberry}, {Eisenstein},
  {Essick}, {Estevez}, {Etienne}, {Etzel}, {Evans}, {Evans}, {Factourovich},
  {Fafone}, {Fair}, {Fairhurst}, {Fan}, {Farinon}, {Farr}, {Farr},
  {Fauchon-Jones}, {Favata}, {Fays}, {Fee}, {Fehrmann}, {Feicht}, {Fejer},
  {Fernandez-Galiana}, {Ferrante}, {Ferreira}, {Ferrini}, {Fidecaro},
  {Finstad}, {Fiori}, {Fiorucci}, {Fishbach}, {Fisher}, {Fitz-Axen},
  {Flaminio}, {Fletcher}, {Fong}, {Font}, {Forsyth}, {Forsyth}, {Fournier},
  {Frasca}, {Frasconi}, {Frei}, {Freise}, {Frey}, {Frey}, {Fries}, {Fritschel},
  {Frolov}, {Fulda}, {Fyffe}, {Gabbard}, {Gadre}, {Gaebel}, {Gair},
  {Gammaitoni}, {Ganija}, {Gaonkar}, {Garcia-Quiros}, {Garufi}, {Gateley},
  {Gaudio}, {Gaur}, {Gayathri}, {Gehrels}, {Gemme}, {Genin}, {Gennai},
  {George}, {George}, {Gergely}, {Germain}, {Ghonge}, {Ghosh}, {Ghosh},
  {Ghosh}, {Giaime}, {Giardina}, {Giazotto}, {Gill}, {Glover}, {Goetz},
  {Goetz}, {Gomes}, {Goncharov}, {Gonz{\'a}lez}, {Gonzalez Castro},
  {Gopakumar}, {Gorodetsky}, {Gossan}, {Gosselin}, {Gouaty}, {Grado}, {Graef},
  {Granata}, {Grant}, {Gras}, {Gray}, {Greco}, {Green}, {Gretarsson}, {Groot},
  {Grote}, {Grunewald}, {Gruning}, {Guidi}, {Guo}, {Gupta}, {Gupta}, {Gushwa},
  {Gustafson}, {Gustafson}, {Halim}, {Hall}, {Hall}, {Hamilton}, {Hammond},
  {Haney}, {Hanke}, {Hanks}, {Hanna}, {Hannam}, {Hannuksela}, {Hanson},
  {Hardwick}, {Harms}, {Harry}, {Harry}, {Hart}, {Haster}, {Haughian}, {Healy},
  {Heidmann}, {Heintze}, {Heitmann}, {Hello}, {Hemming}, {Hendry}, {Heng},
  {Hennig}, {Heptonstall}, {Heurs}, {Hild}, {Hinderer}, {Ho}, {Hoak}, {Hofman},
  {Holt}, {Holz}, {Hopkins}, {Horst}, {Hough}, {Houston}, {Howell}, {Hreibi},
  {Hu}, {Huerta}, {Huet}, {Hughey}, {Husa}, {Huttner}, {Huynh-Dinh}, {Indik},
  {Inta}, {Intini}, {Isa}, {Isac}, {Isi}, {Iyer}, {Izumi}, {Jacqmin}, {Jani},
  {Jaranowski}, {Jawahar}, {Jim{\'e}nez-Forteza}, {Johnson},
  {Johnson-McDaniel}, {Jones}, {Jones}, {Jonker}, {Ju}, {Junker}, {Kalaghatgi},
  {Kalogera}, {Kamai}, {Kandhasamy}, {Kang}, {Kanner}, {Kapadia}, {Karki},
  {Karvinen}, {Kasprzack}, {Kastaun}, {Katolik}, {Katsavounidis}, {Katzman},
  {Kaufer}, {Kawabe}, {K{\'e}f{\'e}lian}, {Keitel}, {Kemball}, {Kennedy},
  {Kent}, {Key}, {Khalili}, {Khan}, {Khan}, {Khan}, {Khazanov}, {Kijbunchoo},
  {Kim}, {Kim}, {Kim}, {Kim}, {Kim}, {Kim}, {Kimbrell}, {King}, {King},
  {Kinley-Hanlon}, {Kirchhoff}, {Kissel}, {Kleybolte}, {Klimenko}, {Knowles},
  {Koch}, {Koehlenbeck}, {Koley}, {Kondrashov}, {Kontos}, {Korobko}, {Korth},
  {Kowalska}, {Kozak}, {Kr{\"a}mer}, {Kringel}, {Krishnan}, {Kr{\'o}lak},
  {Kuehn}, {Kumar}, {Kumar}, {Kumar}, {Kuo}, {Kutynia}, {Kwang}, {Lackey},
  {Lai}, {Landry}, {Lang}, {Lange}, {Lantz}, {Lanza}, {Larson},
  {Lartaux-Vollard}, {Lasky}, {Laxen}, {Lazzarini}, {Lazzaro}, {Leaci},
  {Leavey}, {Lee}, {Lee}, {Lee}, {Lee}, {Lee}, {Lehmann}, {Lenon}, {Leon},
  {Leonardi}, {Leroy}, {Letendre}, {Levin}, {Li}, {Linker}, {Littenberg},
  {Liu}, {Liu}, {Lo}, {Lockerbie}, {London}, {Lord}, {Lorenzini}, {Loriette},
  {Lormand}, {Losurdo}, {Lough}, {Lousto}, {Lovelace}, {L{\"u}ck}, {Lumaca},
  {Lundgren}, {Lynch}, {Ma}, {Macas}, {Macfoy}, {Machenschalk}, {MacInnis},
  {Macleod}, {Maga{\~n}a Hernandez}, {Maga{\~n}a-Sandoval}, {Maga{\~n}a
  Zertuche}, {Magee}, {Majorana}, {Maksimovic}, {Man}, {Mandic}, {Mangano},
  {Mansell}, {Manske}, {Mantovani}, {Marchesoni}, {Marion}, {M{\'a}rka},
  {M{\'a}rka}, {Markakis}, {Markosyan}, {Markowitz}, {Maros}, {Marquina},
  {Marsh}, {Martelli}, {Martellini}, {Martin}, {Martin}, {Martynov}, {Marx},
  {Mason}, {Massera}, {Masserot}, {Massinger}, {Masso-Reid}, {Mastrogiovanni},
  {Matas}, {Matichard}, {Matone}, {Mavalvala}, {Mazumder}, {McCarthy},
  {McClelland}, {McCormick}, {McCuller}, {McGuire}, {McIntyre}, {McIver},
  {McManus}, {McNeill}, {McRae}, {McWilliams}, {Meacher}, {Meadors}, {Mehmet},
  {Meidam}, {Mejuto-Villa}, {Melatos}, {Mendell}, {Mercer}, {Merilh},
  {Merzougui}, {Meshkov}, {Messenger}, {Messick}, {Metzdorff}, {Meyers},
  {Miao}, {Michel}, {Middleton}, {Mikhailov}, {Milano}, {Miller}, {Miller},
  {Miller}, {Millhouse}, {Milovich-Goff}, {Minazzoli}, {Minenkov}, {Ming},
  {Mishra}, {Mitra}, {Mitrofanov}, {Mitselmakher}, {Mittleman}, {Moffa},
  {Moggi}, {Mogushi}, {Mohan}, {Mohapatra}, {Molina}, {Montani}, {Moore},
  {Moraru}, {Moreno}, {Morisaki}, {Morriss}, {Mours}, {Mow-Lowry}, {Mueller},
  {Muir}, {Mukherjee}, {Mukherjee}, {Mukherjee}, {Mukund}, {Mullavey}, {Munch},
  {Mu{\~n}iz}, {Muratore}, {Murray}, {Nagar}, {Napier}, {Nardecchia},
  {Naticchioni}, {Nayak}, {Neilson}, {Nelemans}, {Nelson}, {Nery}, {Neunzert},
  {Nevin}, {Newport}, {Newton}, {Ng}, {Nguyen}, {Nguyen}, {Nichols}, {Nielsen},
  {Nissanke}, {Nitz}, {Noack}, {Nocera}, {Nolting}, {North}, {Nuttall},
  {Oberling}, {O'Dea}, {Ogin}, {Oh}, {Oh}, {Ohme}, {Okada}, {Oliver},
  {Oppermann}, {Oram}, {O'Reilly}, {Ormiston}, {Ortega}, {O'Shaughnessy},
  {Ossokine}, {Ottaway}, {Overmier}, {Owen}, {Pace}, {Page}, {Page}, {Pai},
  {Pai}, {Palamos}, {Palashov}, {Palomba}, {Pal-Singh}, {Pan}, {Pan}, {Pang},
  {Pang}, {Pankow}, {Pannarale}, {Pant}, {Paoletti}, {Paoli}, {Papa}, {Parida},
  {Parker}, {Pascucci}, {Pasqualetti}, {Passaquieti}, {Passuello}, {Patil},
  {Patricelli}, {Pearlstone}, {Pedraza}, {Pedurand}, {Pekowsky}, {Pele},
  {Penn}, {Perez}, {Perreca}, {Perri}, {Pfeiffer}, {Phelps}, {Piccinni},
  {Pichot}, {Piergiovanni}, {Pierro}, {Pillant}, {Pinard}, {Pinto}, {Pirello},
  {Pitkin}, {Poe}, {Poggiani}, {Popolizio}, {Porter}, {Post}, {Powell},
  {Prasad}, {Pratt}, {Pratten}, {Predoi}, {Prestegard}, {Prijatelj},
  {Principe}, {Privitera}, {Prix}, {Prodi}, {Prokhorov}, {Puncken}, {Punturo},
  {Puppo}, {P{\"u}rrer}, {Qi}, {Quetschke}, {Quintero}, {Quitzow-James},
  {Raab}, {Rabeling}, {Radkins}, {Raffai}, {Raja}, {Rajan}, {Rajbhandari},
  {Rakhmanov}, {Ramirez}, {Ramos-Buades}, {Rapagnani}, {Raymond}, {Razzano},
  {Read}, {Regimbau}, {Rei}, {Reid}, {Reitze}, {Ren}, {Reyes}, {Ricci},
  {Ricker}, {Rieger}, {Riles}, {Rizzo}, {Robertson}, {Robie}, {Robinet},
  {Rocchi}, {Rolland}, {Rollins}, {Roma}, {Romano}, {Romano}, {Romel}, {Romie},
  {Rosi{\'n}ska}, {Ross}, {Rowan}, {R{\"u}diger}, {Ruggi}, {Rutins}, {Ryan},
  {Sachdev}, {Sadecki}, {Sadeghian}, {Sakellariadou}, {Salconi}, {Saleem},
  {Salemi}, {Samajdar}, {Sammut}, {Sampson}, {Sanchez}, {Sanchez},
  {Sanchis-Gual}, {Sandberg}, {Sanders}, {Sassolas}, {Sathyaprakash},
  {Saulson}, {Sauter}, {Savage}, {Sawadsky}, {Schale}, {Scheel}, {Scheuer},
  {Schmidt}, {Schmidt}, {Schnabel}, {Schofield}, {Sch{\"o}nbeck}, {Schreiber},
  {Schuette}, {Schulte}, {Schutz}, {Schwalbe}, {Scott}, {Scott}, {Seidel},
  {Sellers}, {Sengupta}, {Sentenac}, {Sequino}, {Sergeev}, {Shaddock},
  {Shaffer}, {Shah}, {Shahriar}, {Shaner}, {Shao}, {Shapiro}, {Shawhan},
  {Sheperd}, {Shoemaker}, {Shoemaker}, {Siellez}, {Siemens}, {Sieniawska},
  {Sigg}, {Silva}, {Singer}, {Singh}, {Singhal}, {Sintes}, {Slagmolen},
  {Smith}, {Smith}, {Smith}, {Somala}, {Son}, {Sonnenberg}, {Sorazu},
  {Sorrentino}, {Souradeep}, {Spencer}, {Srivastava}, {Staats}, {Staley},
  {Steinke}, {Steinlechner}, {Steinlechner}, {Steinmeyer}, {Stevenson},
  {Stone}, {Stops}, {Strain}, {Stratta}, {Strigin}, {Strunk}, {Sturani},
  {Stuver}, {Summerscales}, {Sun}, {Sunil}, {Suresh}, {Sutton}, {Swinkels},
  {Szczepa{\'n}czyk}, {Tacca}, {Tait}, {Talbot}, {Talukder}, {Tanner},
  {T{\'a}pai}, {Taracchini}, {Tasson}, {Taylor}, {Taylor}, {Tewari}, {Theeg},
  {Thies}, {Thomas}, {Thomas}, {Thomas}, {Thorne}, {Thorne}, {Thrane},
  {Tiwari}, {Tiwari}, {Tokmakov}, {Toland}, {Tonelli}, {Tornasi},
  {Torres-Forn{\'e}}, {Torrie}, {T{\"o}yr{\"a}}, {Travasso}, {Traylor},
  {Trinastic}, {Tringali}, {Trozzo}, {Tsang}, {Tse}, {Tso}, {Tsukada}, {Tsuna},
  {Tuyenbayev}, {Ueno}, {Ugolini}, {Unnikrishnan}, {Urban}, {Usman},
  {Vahlbruch}, {Vajente}, {Valdes}, {Vallisneri}, {van Bakel}, {van Beuzekom},
  {van den Brand}, {Van Den Broeck}, {Vander-Hyde}, {van der Schaaf}, {van
  Heijningen}, {van Veggel}, {Vardaro}, {Varma}, {Vass}, {Vas{\'u}th},
  {Vecchio}, {Vedovato}, {Veitch}, {Veitch}, {Venkateswara}, {Venugopalan},
  {Verkindt}, {Vetrano}, {Vicer{\'e}}, {Viets}, {Vinciguerra}, {Vine}, {Vinet},
  {Vitale}, {Vo}, {Vocca}, {Vorvick}, {Vyatchanin}, {Wade}, {Wade}, {Wade},
  {Walet}, {Walker}, {Wallace}, {Walsh}, {Wang}, {Wang}, {Wang}, {Wang},
  {Wang}, {Ward}, {Warner}, {Was}, {Watchi}, {Weaver}, {Wei}, {Weinert},
  {Weinstein}, {Weiss}, {Wen}, {Wessel}, {We{\ss}els}, {Westerweck},
  {Westphal}, {Wette}, {Whelan}, {Whitcomb}, {Whiting}, {Whittle}, {Wilken},
  {Williams}, {Williams}, {Williamson}, {Willis}, {Willke}, {Wimmer},
  {Winkler}, {Wipf}, {Wittel}, {Woan}, {Woehler}, {Wofford}, {Wong}, {Worden},
  {Wright}, {Wu}, {Wysocki}, {Xiao}, {Yamamoto}, {Yancey}, {Yang}, {Yap},
  {Yazback}, {Yu}, {Yu}, {Yvert}, {Zadro{\.Z}ny}, {Zanolin}, {Zelenova},
  {Zendri}, {Zevin}, {Zhang}, {Zhang}, {Zhang}, {Zhang}, {Zhao}, {Zhou},
  {Zhou}, {Zhu}, {Zhu}, {Zimmerman}, {Zucker}, {Zweizig}, {LIGO Scientific
  Collaboration}, \& {Virgo Collaboration}}]{2017PhRvL.119p1101A}
{Abbott}, B.~P., {Abbott}, R., {Abbott}, T.~D., {et~al.} 2017, \prl, 119,
  161101, \dodoi{10.1103/PhysRevLett.119.161101}

\bibitem[{{Abbott} {et~al.}(2018){Abbott}, {Abbott}, {Abbott}, {Abernathy},
  {Acernese}, {Ackley}, {Adams}, {Adams}, {Addesso}, {Adhikari}, {Adya},
  {Affeldt}, {Agathos}, {Agatsuma}, {Aggarwal}, {Aguiar}, {Aiello}, {Ain},
  {Ajith}, {Akutsu}, {Allen}, {Allocca}, {Altin}, {Ananyeva}, {Anderson},
  {Anderson}, {Ando}, {Appert}, {Arai}, {Araya}, {Araya}, {Areeda}, {Arnaud},
  {Arun}, {Asada}, {Ascenzi}, {Ashton}, {Aso}, {Ast}, {Aston}, {Astone},
  {Atsuta}, {Aufmuth}, {Aulbert}, {Avila-Alvarez}, {Awai}, {Babak}, {Bacon},
  {Bader}, {Baiotti}, {Baker}, {Baldaccini}, {Ballardin}, {Ballmer},
  {Barayoga}, {Barclay}, {Barish}, {Barker}, {Barone}, {Barr}, {Barsotti},
  {Barsuglia}, {Barta}, {Bartlett}, {Barton}, {Bartos}, {Bassiri}, {Basti},
  {Batch}, {Baune}, {Bavigadda}, {Bazzan}, {B{\'e}csy}, {Beer}, {Bejger},
  {Belahcene}, {Belgin}, {Bell}, {Berger}, {Bergmann}, {Berry}, {Bersanetti},
  {Bertolini}, {Betzwieser}, {Bhagwat}, {Bhandare}, {Bilenko}, {Billingsley},
  {Billman}, {Birch}, {Birney}, {Birnholtz}, {Biscans}, {Bisht}, {Bitossi},
  {Biwer}, {Bizouard}, {Blackburn}, {Blackman}, {Blair}, {Blair}, {Blair},
  {Bloemen}, {Bock}, {Boer}, {Bogaert}, {Bohe}, {Bondu}, {Bonnand}, {Boom},
  {Bork}, {Boschi}, {Bose}, {Bouffanais}, {Bozzi}, {Bradaschia}, {Brady},
  {Braginsky}, {Branchesi}, {Brau}, {Briant}, {Brillet}, {Brinkmann},
  {Brisson}, {Brockill}, {Broida}, {Brooks}, {Brown}, {Brown}, {Brown},
  {Brunett}, {Buchanan}, {Buikema}, {Bulik}, {Bulten}, {Buonanno}, {Buskulic},
  {Buy}, {Byer}, {Cabero}, {Cadonati}, {Cagnoli}, {Cahillane}, {Calder{\'o}n
  Bustillo}, {Callister}, {Calloni}, {Camp}, {Cannon}, {Cao}, {Cao}, {Capano},
  {Capocasa}, {Carbognani}, {Caride}, {Casanueva Diaz}, {Casentini}, {Caudill},
  {Cavagli{\`a}}, {Cavalier}, {Cavalieri}, {Cella}, {Cepeda}, {Cerboni
  Baiardi}, {Cerretani}, {Cesarini}, {Chamberlin}, {Chan}, {Chao}, {Charlton},
  {Chassande-Mottin}, {Cheeseboro}, {Chen}, {Chen}, {Cheng}, {Chincarini},
  {Chiummo}, {Chmiel}, {Cho}, {Cho}, {Chow}, {Christensen}, {Chu}, {Chua},
  {Chua}, {Chung}, {Ciani}, {Clara}, {Clark}, {Cleva}, {Cocchieri}, {Coccia},
  {Cohadon}, {Colla}, {Collette}, {Cominsky}, {Constancio}, {Conti}, {Cooper},
  {Corbitt}, {Cornish}, {Corsi}, {Cortese}, {Costa}, {Coughlin}, {Coughlin},
  {Coulon}, {Countryman}, {Couvares}, {Covas}, {Cowan}, {Coward}, {Cowart},
  {Coyne}, {Coyne}, {Creighton}, {Creighton}, {Cripe}, {Crowder}, {Cullen},
  {Cumming}, {Cunningham}, {Cuoco}, {Dal Canton}, {Danilishin}, {D'Antonio},
  {Danzmann}, {Dasgupta}, {da Silva Costa}, {Dattilo}, {Dave}, {Davier},
  {Davies}, {Davis}, {Daw}, {Day}, {Day}, {de}, {Debra}, {Debreczeni},
  {Degallaix}, {de Laurentis}, {Del{\'e}glise}, {Del Pozzo}, {Denker}, {Dent},
  {Dergachev}, {De Rosa}, {Derosa}, {Desalvo}, {Devine}, {Dhurandhar},
  {D{\'\i}az}, {di Fiore}, {di Giovanni}, {di Girolamo}, {di Lieto}, {di Pace},
  {di Palma}, {di Virgilio}, {Doctor}, {Doi}, {Dolique}, {Donovan}, {Dooley},
  {Doravari}, {Dorrington}, {Douglas}, {Dovale {\'A}lvarez}, {Downes}, {Drago},
  {Drever}, {Driggers}, {Du}, {Ducrot}, {Dwyer}, {Eda}, {Edo}, {Edwards},
  {Effler}, {Eggenstein}, {Ehrens}, {Eichholz}, {Eikenberry}, {Eisenstein},
  {Essick}, {Etienne}, {Etzel}, {Evans}, {Evans}, {Everett}, {Factourovich},
  {Fafone}, {Fair}, {Fairhurst}, {Fan}, {Farinon}, {Farr}, {Farr},
  {Fauchon-Jones}, {Favata}, {Fays}, {Fehrmann}, {Fejer}, {Fern{\'a}ndez
  Galiana}, {Ferrante}, {Ferreira}, {Ferrini}, {Fidecaro}, {Fiori}, {Fiorucci},
  {Fisher}, {Flaminio}, {Fletcher}, {Fong}, {Forsyth}, {Fournier}, {Frasca},
  {Frasconi}, {Frei}, {Freise}, {Frey}, {Frey}, {Fries}, {Fritschel}, {Frolov},
  {Fujii}, {Fujimoto}, {Fulda}, {Fyffe}, {Gabbard}, {Gadre}, {Gaebel}, {Gair},
  {Gammaitoni}, {Gaonkar}, {Garufi}, {Gaur}, {Gayathri}, {Gehrels}, {Gemme},
  {Genin}, {Gennai}, {George}, {Gergely}, {Germain}, {Ghonge}, {Ghosh},
  {Ghosh}, {Ghosh}, {Giaime}, {Giardina}, {Giazotto}, {Gill}, {Glaefke},
  {Goetz}, {Goetz}, {Gondan}, {Gonz{\'a}lez}, {Gonzalez Castro}, {Gopakumar},
  {Gorodetsky}, {Gossan}, {Gosselin}, {Gouaty}, {Grado}, {Graef}, {Granata},
  {Grant}, {Gras}, {Gray}, {Greco}, {Green}, {Groot}, {Grote}, {Grunewald},
  {Guidi}, {Guo}, {Gupta}, {Gupta}, {Gushwa}, {Gustafson}, {Gustafson},
  {Hacker}, {Hagiwara}, {Hall}, {Hall}, {Hammond}, {Haney}, {Hanke}, {Hanks},
  {Hanna}, {Hannam}, {Hanson}, {Hardwick}, {Harms}, {Harry}, {Harry}, {Hart},
  {Hartman}, {Haster}, {Haughian}, {Hayama}, {Healy}, {Heidmann}, {Heintze},
  {Heitmann}, {Hello}, {Hemming}, {Hendry}, {Heng}, {Hennig}, {Henry},
  {Heptonstall}, {Heurs}, {Hild}, {Hirose}, {Hoak}, {Hofman}, {Holt}, {Holz},
  {Hopkins}, {Hough}, {Houston}, {Howell}, {Hu}, {Huerta}, {Huet}, {Hughey},
  {Husa}, {Huttner}, {Huynh-Dinh}, {Indik}, {Ingram}, {Inta}, {Ioka}, {Isa},
  {Isac}, {Isi}, {Isogai}, {Itoh}, {Iyer}, {Izumi}, {Jacqmin}, {Jani},
  {Jaranowski}, {Jawahar}, {Jim{\'e}nez-Forteza}, {Johnson}, {Jones}, {Jones},
  {Jonker}, {Ju}, {Junker}, {Kagawa}, {Kajita}, {Kakizaki}, {Kalaghatgi},
  {Kalogera}, {Kamiizumi}, {Kanda}, {Kandhasamy}, {Kanemura}, {Kaneyama},
  {Kang}, {Kanner}, {Karki}, {Karvinen}, {Kasprzack}, {Kataoka},
  {Katsavounidis}, {Katzman}, {Kaufer}, {Kaur}, {Kawabe}, {Kawai}, {Kawamura},
  {K{\'e}f{\'e}lian}, {Keitel}, {Kelley}, {Kennedy}, {Key}, {Khalili}, {Khan},
  {Khan}, {Khan}, {Khazanov}, {Kijbunchoo}, {Kim}, {Kim}, {Kim}, {Kim}, {Kim},
  {Kim}, {Kimbrell}, {Kimura}, {King}, {King}, {Kirchhoff}, {Kissel}, {Klein},
  {Kleybolte}, {Klimenko}, {Koch}, {Koehlenbeck}, {Kojima}, {Kokeyama},
  {Koley}, {Komori}, {Kondrashov}, {Kontos}, {Korobko}, {Korth}, {Kotake},
  {Kowalska}, {Kozak}, {Kr{\"a}mer}, {Kringel}, {Krishnan}, {Kr{\'o}lak},
  {Kuehn}, {Kumar}, {Kumar}, {Kumar}, {Kuo}, {Kuroda}, {Kutynia}, {Kuwahara},
  {Lackey}, {Landry}, {Lang}, {Lange}, {Lantz}, {Lanza}, {Lartaux-Vollard},
  {Lasky}, {Laxen}, {Lazzarini}, {Lazzaro}, {Leaci}, {Leavey}, {Lebigot},
  {Lee}, {Lee}, {Lee}, {Lee}, {Lee}, {Lehmann}, {Lenon}, {Leonardi}, {Leong},
  {Leroy}, {Letendre}, {Levin}, {Li}, {Libson}, {Littenberg}, {Liu},
  {Lockerbie}, {Lombardi}, {London}, {Lord}, {Lorenzini}, {Loriette},
  {Lormand}, {Losurdo}, {Lough}, {Lousto}, {Lovelace}, {L{\"u}ck}, {Lundgren},
  {Lynch}, {Ma}, {Macfoy}, {Machenschalk}, {Macinnis}, {MacLeod},
  {Maga{\~n}a-Sandoval}, {Majorana}, {Maksimovic}, {Malvezzi}, {Man}, {Mandic},
  {Mangano}, {Mano}, {Mansell}, {Manske}, {Mantovani}, {Marchesoni}, {Marchio},
  {Marion}, {M{\'a}rka}, {M{\'a}rka}, {Markosyan}, {Maros}, {Martelli},
  {Martellini}, {Martin}, {Martynov}, {Mason}, {Masserot}, {Massinger},
  {Masso-Reid}, {Mastrogiovanni}, {Matichard}, {Matone}, {Matsumoto},
  {Matsushima}, {Mavalvala}, {Mazumder}, {McCarthy}, {McClelland}, {McCormick},
  {McGrath}, {McGuire}, {McIntyre}, {McIver}, {McManus}, {McRae}, {McWilliams},
  {Meacher}, {Meadors}, {Meidam}, {Melatos}, {Mendell}, {Mendoza-Gandara},
  {Mercer}, {Merilh}, {Merzougui}, {Meshkov}, {Messenger}, {Messick},
  {Metzdorff}, {Meyers}, {Mezzani}, {Miao}, {Michel}, {Michimura}, {Middleton},
  {Mikhailov}, {Milano}, {Miller}, {Miller}, {Miller}, {Miller}, {Millhouse},
  {Minenkov}, {Ming}, {Mirshekari}, {Mishra}, {Mitrofanov}, {Mitselmakher},
  {Mittleman}, {Miyakawa}, {Miyamoto}, {Miyamoto}, {Miyoki}, {Moggi}, {Mohan},
  {Mohapatra}, {Montani}, {Moore}, {Moore}, {Moraru}, {Moreno}, {Morii},
  {Morisaki}, {Moriwaki}, {Morriss}, {Mours}, {Mow-Lowry}, {Mueller}, {Muir},
  {Mukherjee}, {Mukherjee}, {Mukherjee}, {Mukund}, {Mullavey}, {Munch},
  {Muniz}, {Murray}, {Mytidis}, {Nagano}, {Nakamura}, {Nakamura}, {Nakano},
  {Nakano}, {Nakano}, {Nakao}, {Napier}, {Nardecchia}, {Narikawa},
  {Naticchioni}, {Nelemans}, {Nelson}, {Neri}, {Nery}, {Neunzert}, {Newport},
  {Newton}, {Nguyen}, {Ni}, {Nielsen}, {Nissanke}, {Nitz}, {Noack}, {Nocera},
  {Nolting}, {Normandin}, {Nuttall}, {Oberling}, {Ochsner}, {Oelker}, {Ogin},
  {Oh}, {Oh}, {Ohashi}, {Ohishi}, {Ohkawa}, {Ohme}, {Okutomi}, {Oliver}, {Ono},
  {Ono}, {Oohara}, {Oppermann}, {Oram}, {O'Reilly}, {O'Shaughnessy}, {Ottaway},
  {Overmier}, {Owen}, {Pace}, {Page}, {Pai}, {Pai}, {Palamos}, {Palashov},
  {Palomba}, {Pal-Singh}, {Pan}, {Pankow}, {Pannarale}, {Pant}, {Paoletti},
  {Paoli}, {Papa}, {Paris}, {Parker}, {Pascucci}, {Pasqualetti}, {Passaquieti},
  {Passuello}, {Patricelli}, {Pearlstone}, {Pedraza}, {Pedurand}, {Pekowsky},
  {Pele}, {Pe{\~n}a Arellano}, {Penn}, {Perez}, {Perreca}, {Perri}, {Pfeiffer},
  {Phelps}, {Piccinni}, {Pichot}, {Piergiovanni}, {Pierro}, {Pillant},
  {Pinard}, {Pinto}, {Pitkin}, {Poe}, {Poggiani}, {Popolizio}, {Post},
  {Powell}, {Prasad}, {Pratt}, {Predoi}, {Prestegard}, {Prijatelj}, {Principe},
  {Privitera}, {Prodi}, {Prokhorov}, {Puncken}, {Punturo}, {Puppo},
  {P{\"u}rrer}, {Qi}, {Qin}, {Qiu}, {Quetschke}, {Quintero}, {Quitzow-James},
  {Raab}, {Rabeling}, {Radkins}, {Raffai}, {Raja}, {Rajan}, {Rakhmanov},
  {Rapagnani}, {Raymond}, {Razzano}, {Re}, {Read}, {Regimbau}, {Rei}, {Reid},
  {Reitze}, {Rew}, {Reyes}, {Rhoades}, {Ricci}, {Riles}, {Rizzo}, {Robertson},
  {Robie}, {Robinet}, {Rocchi}, {Rolland}, {Rollins}, {Roma}, {Romano},
  {Romie}, {Rosi{\'n}ska}, {Rowan}, {R{\"u}diger}, {Ruggi}, {Ryan}, {Sachdev},
  {Sadecki}, {Sadeghian}, {Sago}, {Saijo}, {Saito}, {Sakai}, {Sakellariadou},
  {Salconi}, {Saleem}, {Salemi}, {Samajdar}, {Sammut}, {Sampson}, {Sanchez},
  {Sandberg}, {Sanders}, {Sasaki}, {Sassolas}, {Sathyaprakash}, {Sato}, {Sato},
  {Saulson}, {Sauter}, {Savage}, {Sawadsky}, {Schale}, {Scheuer}, {Schmidt},
  {Schmidt}, {Schmidt}, {Schnabel}, {Schofield}, {Sch{\"o}nbeck}, {Schreiber},
  {Schuette}, {Schutz}, {Schwalbe}, {Scott}, {Scott}, {Sekiguchi}, {Sekiguchi},
  {Sellers}, {Sengupta}, {Sentenac}, {Sequino}, {Sergeev}, {Setyawati},
  {Shaddock}, {Shaffer}, {Shahriar}, {Shapiro}, {Shawhan}, {Sheperd},
  {Shibata}, {Shikano}, {Shimoda}, {Shoda}, {Shoemaker}, {Shoemaker},
  {Siellez}, {Siemens}, {Sieniawska}, {Sigg}, {Silva}, {Singer}, {Singer},
  {Singh}, {Singh}, {Singhal}, {Sintes}, {Slagmolen}, {Smith}, {Smith},
  {Smith}, {Somiya}, {Son}, {Sorazu}, {Sorrentino}, {Souradeep}, {Spencer},
  {Srivastava}, {Staley}, {Steinke}, {Steinlechner}, {Steinlechner},
  {Steinmeyer}, {Stephens}, {Stevenson}, {Stone}, {Strain}, {Straniero},
  {Stratta}, {Strigin}, {Sturani}, {Stuver}, {Sugimoto}, {Summerscales}, {Sun},
  {Sunil}, {Sutton}, {Suzuki}, {Swinkels}, {Szczepa{\'n}czyk}, {Tacca},
  {Tagoshi}, {Takada}, {Takahashi}, {Takahashi}, {Takamori}, {Talukder},
  {Tanaka}, {Tanaka}, {Tanaka}, {Tanner}, {T{\'a}pai}, {Taracchini}, {Tatsumi},
  {Taylor}, {Telada}, {Theeg}, {Thomas}, {Thomas}, {Thomas}, {Thorne},
  {Thrane}, {Tippens}, {Tiwari}, {Tiwari}, {Tokmakov}, {Toland}, {Tomaru},
  {Tomlinson}, {Tonelli}, {Tornasi}, {Torrie}, {T{\"o}yr{\"a}}, {Travasso},
  {Traylor}, {Trifir{\`o}}, {Trinastic}, {Tringali}, {Trozzo}, {Tse}, {Tso},
  {Tsubono}, {Tsuzuki}, {Turconi}, {Tuyenbayev}, {Uchiyama}, {Uehara}, {Ueki},
  {Ueno}, {Ugolini}, {Unnikrishnan}, {Urban}, {Ushiba}, {Usman}, {Vahlbruch},
  {Vajente}, {Valdes}, {van Bakel}, {van Beuzekom}, {van den Brand}, {van den
  Broeck}, {Vander-Hyde}, {van der Schaaf}, {van Heijningen}, {van Putten},
  {van Veggel}, {Vardaro}, {Varma}, {Vass}, {Vas{\'u}th}, {Vecchio},
  {Vedovato}, {Veitch}, {Veitch}, {Venkateswara}, {Venugopalan}, {Verkindt},
  {Vetrano}, {Vicer{\'e}}, {Viets}, {Vinciguerra}, {Vine}, {Vinet}, {Vitale},
  {Vo}, {Vocca}, {Vorvick}, {Voss}, {Vousden}, {Vyatchanin}, {Wade}, {Wade},
  {Wade}, {Wakamatsu}, {Walker}, {Wallace}, {Walsh}, {Wang}, {Wang}, {Wang},
  {Wang}, {Ward}, {Warner}, {Was}, {Watchi}, {Weaver}, {Wei}, {Weinert},
  {Weinstein}, {Weiss}, {Wen}, {We{\ss}els}, {Westphal}, {Wette}, {Whelan},
  {Whiting}, {Whittle}, {Williams}, {Williams}, {Williamson}, {Willis},
  {Willke}, {Wimmer}, {Winkler}, {Wipf}, {Wittel}, {Woan}, {Woehler}, {Worden},
  {Wright}, {Wu}, {Wu}, {Yam}, {Yamamoto}, {Yamamoto}, {Yamamoto}, {Yancey},
  {Yano}, {Yap}, {Yokoyama}, {Yokozawa}, {Yoon}, {Yu}, {Yu}, {Yuzurihara},
  {Yvert}, {Zadro{\.z}ny}, {Zangrando}, {Zanolin}, {Zeidler}, {Zendri},
  {Zevin}, {Zhang}, {Zhang}, {Zhang}, {Zhang}, {Zhao}, {Zhou}, {Zhou}, {Zhu},
  {Zhu}, {Zucker}, {Zweizig}, {Kagra Collaboration}, \& {VIRGO
  Collaboration}}]{2018LRR....21....3A}
---. 2018, Living Reviews in Relativity, 21, 3,
  \dodoi{10.1007/s41114-018-0012-9}

\bibitem[{Acernese {et~al.}(2014)Acernese, Agathos, Agatsuma, Aisa, Allemandou,
  Allocca, Amarni, Astone, Balestri, Ballardin, Barone, Baronick, Barsuglia,
  Basti, Basti, Bauer, Bavigadda, Bejger, Beker, Belczynski, Bersanetti,
  Bertolini, Bitossi, Bizouard, Bloemen, Blom, Boer, Bogaert, Bondi, Bondu,
  Bonelli, Bonnand, Boschi, Bosi, Bouedo, Bradaschia, Branchesi, Briant,
  Brillet, Brisson, Bulik, Bulten, Buskulic, Buy, Cagnoli, Calloni, Campeggi,
  Canuel, Carbognani, Cavalier, Cavalieri, Cella, Cesarini, Chassande-Mottin,
  Chincarini, Chiummo, Chua, Cleva, Coccia, Cohadon, Colla, Colombini, Conte,
  Coulon, Cuoco, Dalmaz, D'Antonio, Dattilo, Davier, Day, Debreczeni,
  Degallaix, Del{\'e}glise, Pozzo, Dereli, Rosa, Fiore, Lieto, Virgilio, Doets,
  Dolique, Drago, Ducrot, Endr{\H o}czi, Fafone, Farinon, Ferrante, Ferrini,
  Fidecaro, Fiori, Flaminio, Fournier, Franco, Frasca, Frasconi, Gammaitoni,
  Garufi, Gaspard, Gatto, Gemme, Gendre, Genin, Gennai, Ghosh, Giacobone,
  Giazotto, Gouaty, Granata, Greco, Groot, Guidi, Harms, Heidmann, Heitmann,
  Hello, Hemming, Hennes, Hofman, Jaranowski, Jonker, Kasprzack,
  K{\'e}f{\'e}lian, Kowalska, Kraan, Kr{\'o}lak, Kutynia, Lazzaro, Leonardi,
  Leroy, Letendre, Li, Lieunard, Lorenzini, Loriette, Losurdo, Magazz{\`u},
  Majorana, Maksimovic, Malvezzi, Man, Mangano, Mantovani, Marchesoni, Marion,
  Marque, Martelli, Martellini, Masserot, Meacher, Meidam, Mezzani, Michel,
  Milano, Minenkov, Moggi, Mohan, Montani, Morgado, Mours, Mul, Nagy,
  Nardecchia, Naticchioni, Nelemans, Neri, Neri, Nocera, Pacaud, Palomba,
  Paoletti, Paoli, Pasqualetti, Passaquieti, Passuello, Perciballi, Petit,
  Pichot, Piergiovanni, Pillant, Piluso, Pinard, Poggiani, Prijatelj, Prodi,
  Punturo, Puppo, Rabeling, R{\'a}cz, Rapagnani, Razzano, Re, Regimbau, Ricci,
  Robinet, Rocchi, Rolland, Romano, Rosi{\'n}ska, Ruggi, Saracco, Sassolas,
  Schimmel, Sentenac, Sequino, Shah, Siellez, Straniero, Swinkels, Tacca,
  Tonelli, Travasso, Turconi, Vajente, van Bakel, van Beuzekom, van~den Brand,
  Broeck, van~der Sluys, van Heijningen, Vas{\'u}th, Vedovato, Veitch,
  Verkindt, Vetrano, Vicer{\'e}, Vinet, Visser, Vocca, Ward, Was, Wei, Yvert,
  {\.z}ny, \& Zendri}]{Virgo_detector}
Acernese, F., Agathos, M., Agatsuma, K., {et~al.} 2014, Classical and Quantum
  Gravity, 32, 024001, \dodoi{10.1088/0264-9381/32/2/024001}

\bibitem[{{Ahumada} {et~al.}(2022){Ahumada}, {Anand}, {Coughlin}, {Andreoni},
  {Kool}, {Kumar}, {Reusch}, {Sagu{\'e}s-Carracedo}, {Stein}, {Cenko},
  {Kasliwal}, {Singer}, {Dunwoody}, {Mangan}, {Bhalerao}, {Bulla}, {Burns},
  {Graham}, {Kaplan}, {Perley}, {Almualla}, {Bloom}, {Cunningham}, {De},
  {Gatkine}, {Ho}, {Karambelkar}, {Kong}, {Yao}, {Anupama}, {Barway}, {Ghosh},
  {Itoh}, {McBreen}, {Bellm}, {Fremling}, {Laher}, {Mahabal}, {Riddle},
  {Rosnet}, {Rusholme}, {Smith}, {Sollerman}, {Bissaldi}, {Fletcher},
  {Hamburg}, {Mailyan}, {Malacaria}, \& {Roberts}}]{2022ApJ...932...40A}
{Ahumada}, T., {Anand}, S., {Coughlin}, M.~W., {et~al.} 2022, \apj, 932, 40,
  \dodoi{10.3847/1538-4357/ac6c29}

\bibitem[{Akutsu {et~al.}(2021)Akutsu, Ando, Arai, Arai, Araki, Araya, Aritomi,
  Asada, Aso, Bae, Bae, Baiotti, Bajpai, Barton, Cannon, Cao, Capocasa, Chan,
  Chen, Chen, Chen, Chiang, Chu, Chu, Eguchi, Enomoto, Flaminio, Fujii,
  Fujikawa, Fukunaga, Fukushima, Gao, Ge, Ha, Hagiwara, Haino, Han, Hasegawa,
  Hattori, Hayakawa, Hayama, Himemoto, Hiranuma, Hirata, Hirose, Hong, Hsieh,
  Huang, Huang, Huang, Huang, Huang, Hui, Ide, Ikenoue, Imam, Inayoshi, Inoue,
  Ioka, Ito, Itoh, Izumi, Jeon, Jin, Jung, Jung, Kaihotsu, Kajita, Kakizaki,
  Kamiizumi, Kanda, Kang, Kawaguchi, Kawai, Kawasaki, Kim, Kim, Kim, Kim, Kim,
  Kimura, Kita, Kitazawa, Kojima, Kokeyama, Komori, Kong, Kotake, Kozakai,
  Kozu, Kumar, Kume, Kuo, Kuo, Kuromiya, Kuroyanagi, Kusayanagi, Kwak, Lee,
  Lee, Lee, Leonardi, Li, Lin, Lin, Lin, Lin, Lin, Liu, Luo, Majorana, Marchio,
  Michimura, Mio, Miyakawa, Miyamoto, Miyazaki, Miyo, Miyoki, Mori, Morisaki,
  Moriwaki, Nagano, Nagano, Nakamura, Nakano, Nakano, Nakashima, Nakayama,
  Narikawa, Naticchioni, Negishi, Nguyen~Quynh, Ni, Nishizawa, Nozaki, Obuchi,
  Ogaki, Oh, Oh, Oh, Ohashi, Ohishi, Ohkawa, Ohta, Okutani, Okutomi, Oohara,
  Ooi, Oshino, Otabe, Pan, Pang, Parisi, Park, Pe{\~n}a~Arellano, Pinto, Sago,
  Saito, Saito, Sakai, Sakai, Sakuno, Sato, Sato, Sawada, Sekiguchi, Sekiguchi,
  Shao, Shibagaki, Shimizu, Shimoda, Shimode, Shinkai, Shishido, Shoda, Somiya,
  Son, Sotani, Sugimoto, Suresh, Suzuki, Suzuki, Tagoshi, Takahashi, Takahashi,
  Takamori, Takano, Takeda, Takeda, Tanaka, Tanaka, Tanaka, Tanaka, Tanaka,
  Tanioka, Tapia San~Martin, Telada, Tomaru, Tomigami, Tomura, Travasso,
  Trozzo, Tsang, Tsao, Tsubono, Tsuchida, Tsutsui, Tsuzuki, Tuyenbayev,
  Uchikata, Uchiyama, Ueda, Uehara, Ueno, Ueshima, Uraguchi, Ushiba, van
  Putten, Vocca, Wang, Washimi, Wu, Wu, Wu, Xu, Yamada, Yamamoto, Yamamoto,
  Yamamoto, Yamashita, Yamazaki, Yang, Yokogawa, Yokoyama, Yokozawa, Yoshioka,
  Yuzurihara, Zeidler, Zhan, Zhang, Zhao, \& Zhu}]{KAGRA_detector}
Akutsu, T., Ando, M., Arai, K., {et~al.} 2021, Progress of Theoretical and
  Experimental Physics, 2021, \dodoi{10.1093/ptep/ptab018}

\bibitem[{{Anand} {et~al.}(2021){Anand}, {Coughlin}, {Kasliwal}, {Bulla},
  {Ahumada}, {Sagu{\'e}s Carracedo}, {Almualla}, {Andreoni}, {Stein},
  {Foucart}, {Singer}, {Sollerman}, {Bellm}, {Bolin}, {Caballero-Garc{\'\i}a},
  {Castro-Tirado}, {Cenko}, {De}, {Dekany}, {Duev}, {Feeney}, {Fremling},
  {Goldstein}, {Golkhou}, {Graham}, {Guessoum}, {Hankins}, {Hu}, {Kong},
  {Kool}, {Kulkarni}, {Kumar}, {Laher}, {Masci}, {Mr{\'o}z}, {Nissanke},
  {Porter}, {Reusch}, {Riddle}, {Rosnet}, {Rusholme}, {Serabyn},
  {S{\'a}nchez-Ram{\'\i}rez}, {Rigault}, {Shupe}, {Smith}, {Soumagnac},
  {Walters}, \& {Valeev}}]{2021NatAs...5...46A}
{Anand}, S., {Coughlin}, M.~W., {Kasliwal}, M.~M., {et~al.} 2021, Nature
  Astronomy, 5, 46, \dodoi{10.1038/s41550-020-1183-3}

\bibitem[{{Andreoni} {et~al.}(2020){Andreoni}, {Kool}, {Sagu{\'e}s Carracedo},
  {Kasliwal}, {Bulla}, {Ahumada}, {Coughlin}, {Anand}, {Sollerman}, {Goobar},
  {Kaplan}, {Loveridge}, {Karambelkar}, {Cooke}, {Bagdasaryan}, {Bellm},
  {Cenko}, {Cook}, {De}, {Dekany}, {Delacroix}, {Drake}, {Duev}, {Fremling},
  {Golkhou}, {Graham}, {Hale}, {Kulkarni}, {Kupfer}, {Laher}, {Mahabal},
  {Masci}, {Rusholme}, {Smith}, {Tzanidakis}, {Van Sistine}, \&
  {Yao}}]{2020ApJ...904..155A}
{Andreoni}, I., {Kool}, E.~C., {Sagu{\'e}s Carracedo}, A., {et~al.} 2020, \apj,
  904, 155, \dodoi{10.3847/1538-4357/abbf4c}

\bibitem[{{Andreoni} {et~al.}(2021){Andreoni}, {Coughlin}, {Kool}, {Kasliwal},
  {Kumar}, {Bhalerao}, {Carracedo}, {Ho}, {Pang}, {Saraogi}, {Sharma},
  {Shenoy}, {Burns}, {Ahumada}, {Anand}, {Singer}, {Perley}, {De}, {Fremling},
  {Bellm}, {Bulla}, {Crellin-Quick}, {Dietrich}, {Drake}, {Duev}, {Goobar},
  {Graham}, {Kaplan}, {Kulkarni}, {Laher}, {Mahabal}, {Shupe}, {Sollerman},
  {Walters}, \& {Yao}}]{2021ApJ...918...63A}
{Andreoni}, I., {Coughlin}, M.~W., {Kool}, E.~C., {et~al.} 2021, \apj, 918, 63,
  \dodoi{10.3847/1538-4357/ac0bc7}

\bibitem[{{Antier} {et~al.}(2020){Antier}, {Agayeva}, {Almualla}, {Awiphan},
  {Baransky}, {Barynova}, {Beradze}, {Bla{\v{z}}ek}, {Bo{\"e}r}, {Burkhonov},
  {Christensen}, {Coleiro}, {Corre}, {Coughlin}, {Crisp}, {Dietrich}, {Ducoin},
  {Duverne}, {Marchal-Duval}, {Gendre}, {Gokuldass}, {Eggenstein}, {Eymar},
  {Hello}, {Howell}, {Ismailov}, {Kann}, {Karpov}, {Klotz}, {Kochiashvili},
  {Lachaud}, {Leroy}, {Lin}, {Li}, {Ma{\v{s}}ek}, {Mo}, {Menard}, {Morris},
  {Noysena}, {Orange}, {Prouza}, {Rattanamala}, {Sadibekova}, {Saint-Gelais},
  {Serrau}, {Simon}, {Stachie}, {Th{\"o}ne}, {Tillayev}, {Turpin}, {Postigo},
  {Vasylenko}, {Vidadi}, {Was}, {Wang}, {Zhang}, {Zhang}, \&
  {Zhang}}]{2020MNRAS.497.5518A}
{Antier}, S., {Agayeva}, S., {Almualla}, M., {et~al.} 2020, \mnras, 497, 5518,
  \dodoi{10.1093/mnras/staa1846}

\bibitem[{{Astropy Collaboration} {et~al.}(2013){Astropy Collaboration},
  {Robitaille}, {Tollerud}, {Greenfield}, {Droettboom}, {Bray}, {Aldcroft},
  {Davis}, {Ginsburg}, {Price-Whelan}, {Kerzendorf}, {Conley}, {Crighton},
  {Barbary}, {Muna}, {Ferguson}, {Grollier}, {Parikh}, {Nair}, {Unther},
  {Deil}, {Woillez}, {Conseil}, {Kramer}, {Turner}, {Singer}, {Fox}, {Weaver},
  {Zabalza}, {Edwards}, {Azalee Bostroem}, {Burke}, {Casey}, {Crawford},
  {Dencheva}, {Ely}, {Jenness}, {Labrie}, {Lim}, {Pierfederici}, {Pontzen},
  {Ptak}, {Refsdal}, {Servillat}, \& {Streicher}}]{astropy:2013}
{Astropy Collaboration}, {Robitaille}, T.~P., {Tollerud}, E.~J., {et~al.} 2013,
  \aap, 558, A33, \dodoi{10.1051/0004-6361/201322068}

\bibitem[{{Astropy Collaboration} {et~al.}(2018){Astropy Collaboration},
  {Price-Whelan}, {Sip{\H{o}}cz}, {G{\"u}nther}, {Lim}, {Crawford}, {Conseil},
  {Shupe}, {Craig}, {Dencheva}, {Ginsburg}, {Vand erPlas}, {Bradley},
  {P{\'e}rez-Su{\'a}rez}, {de Val-Borro}, {Aldcroft}, {Cruz}, {Robitaille},
  {Tollerud}, {Ardelean}, {Babej}, {Bach}, {Bachetti}, {Bakanov}, {Bamford},
  {Barentsen}, {Barmby}, {Baumbach}, {Berry}, {Biscani}, {Boquien}, {Bostroem},
  {Bouma}, {Brammer}, {Bray}, {Breytenbach}, {Buddelmeijer}, {Burke},
  {Calderone}, {Cano Rodr{\'\i}guez}, {Cara}, {Cardoso}, {Cheedella}, {Copin},
  {Corrales}, {Crichton}, {D'Avella}, {Deil}, {Depagne}, {Dietrich}, {Donath},
  {Droettboom}, {Earl}, {Erben}, {Fabbro}, {Ferreira}, {Finethy}, {Fox},
  {Garrison}, {Gibbons}, {Goldstein}, {Gommers}, {Greco}, {Greenfield},
  {Groener}, {Grollier}, {Hagen}, {Hirst}, {Homeier}, {Horton}, {Hosseinzadeh},
  {Hu}, {Hunkeler}, {Ivezi{\'c}}, {Jain}, {Jenness}, {Kanarek}, {Kendrew},
  {Kern}, {Kerzendorf}, {Khvalko}, {King}, {Kirkby}, {Kulkarni}, {Kumar},
  {Lee}, {Lenz}, {Littlefair}, {Ma}, {Macleod}, {Mastropietro}, {McCully},
  {Montagnac}, {Morris}, {Mueller}, {Mumford}, {Muna}, {Murphy}, {Nelson},
  {Nguyen}, {Ninan}, {N{\"o}the}, {Ogaz}, {Oh}, {Parejko}, {Parley}, {Pascual},
  {Patil}, {Patil}, {Plunkett}, {Prochaska}, {Rastogi}, {Reddy Janga},
  {Sabater}, {Sakurikar}, {Seifert}, {Sherbert}, {Sherwood-Taylor}, {Shih},
  {Sick}, {Silbiger}, {Singanamalla}, {Singer}, {Sladen}, {Sooley},
  {Sornarajah}, {Streicher}, {Teuben}, {Thomas}, {Tremblay}, {Turner},
  {Terr{\'o}n}, {van Kerkwijk}, {de la Vega}, {Watkins}, {Weaver}, {Whitmore},
  {Woillez}, {Zabalza}, \& {Astropy Contributors}}]{astropy:2018}
{Astropy Collaboration}, {Price-Whelan}, A.~M., {Sip{\H{o}}cz}, B.~M., {et~al.}
  2018, \aj, 156, 123, \dodoi{10.3847/1538-3881/aabc4f}

\bibitem[{{Astropy Collaboration} {et~al.}(2022){Astropy Collaboration},
  {Price-Whelan}, {Lim}, {Earl}, {Starkman}, {Bradley}, {Shupe}, {Patil},
  {Corrales}, {Brasseur}, {N{"o}the}, {Donath}, {Tollerud}, {Morris},
  {Ginsburg}, {Vaher}, {Weaver}, {Tocknell}, {Jamieson}, {van Kerkwijk},
  {Robitaille}, {Merry}, {Bachetti}, {G{"u}nther}, {Aldcroft},
  {Alvarado-Montes}, {Archibald}, {B{'o}di}, {Bapat}, {Barentsen}, {Baz{'a}n},
  {Biswas}, {Boquien}, {Burke}, {Cara}, {Cara}, {Conroy}, {Conseil}, {Craig},
  {Cross}, {Cruz}, {D'Eugenio}, {Dencheva}, {Devillepoix}, {Dietrich},
  {Eigenbrot}, {Erben}, {Ferreira}, {Foreman-Mackey}, {Fox}, {Freij}, {Garg},
  {Geda}, {Glattly}, {Gondhalekar}, {Gordon}, {Grant}, {Greenfield}, {Groener},
  {Guest}, {Gurovich}, {Handberg}, {Hart}, {Hatfield-Dodds}, {Homeier},
  {Hosseinzadeh}, {Jenness}, {Jones}, {Joseph}, {Kalmbach}, {Karamehmetoglu},
  {Ka{l}uszy{'n}ski}, {Kelley}, {Kern}, {Kerzendorf}, {Koch}, {Kulumani},
  {Lee}, {Ly}, {Ma}, {MacBride}, {Maljaars}, {Muna}, {Murphy}, {Norman},
  {O'Steen}, {Oman}, {Pacifici}, {Pascual}, {Pascual-Granado}, {Patil},
  {Perren}, {Pickering}, {Rastogi}, {Roulston}, {Ryan}, {Rykoff}, {Sabater},
  {Sakurikar}, {Salgado}, {Sanghi}, {Saunders}, {Savchenko}, {Schwardt},
  {Seifert-Eckert}, {Shih}, {Jain}, {Shukla}, {Sick}, {Simpson},
  {Singanamalla}, {Singer}, {Singhal}, {Sinha}, {Sip{H{o}}cz}, {Spitler},
  {Stansby}, {Streicher}, {{{S}}umak}, {Swinbank}, {Taranu}, {Tewary},
  {Tremblay}, {Val-Borro}, {Van Kooten}, {Vasovi{'c}}, {Verma}, {de Miranda
  Cardoso}, {Williams}, {Wilson}, {Winkel}, {Wood-Vasey}, {Xue}, {Yoachim},
  {Zhang}, {Zonca}, \& {Astropy Project Contributors}}]{astropy:2022}
{Astropy Collaboration}, {Price-Whelan}, A.~M., {Lim}, P.~L., {et~al.} 2022,
  apj, 935, 167, \dodoi{10.3847/1538-4357/ac7c74}

\bibitem[{{Astudillo} {et~al.}(2020){Astudillo}, {Protopapas}, {Pichara}, \&
  {Huijse}}]{2020AJ....159...16A}
{Astudillo}, J., {Protopapas}, P., {Pichara}, K., \& {Huijse}, P. 2020, \aj,
  159, 16, \dodoi{10.3847/1538-3881/ab557d}

\bibitem[{{Bauswein} {et~al.}(2017){Bauswein}, {Just}, {Janka}, \&
  {Stergioulas}}]{2017ApJ...850L..34B}
{Bauswein}, A., {Just}, O., {Janka}, H.-T., \& {Stergioulas}, N. 2017, \apjl,
  850, L34, \dodoi{10.3847/2041-8213/aa9994}

\bibitem[{{Bellm}(2014)}]{2014htu..conf...27B}
{Bellm}, E. 2014, in The Third Hot-wiring the Transient Universe Workshop, ed.
  P.~R. {Wozniak}, M.~J. {Graham}, A.~A. {Mahabal}, \& R.~{Seaman}, 27--33.
\newblock \doarXiv{1410.8185}

\bibitem[{{Bellm} \& {Kulkarni}(2017)}]{2017NatAs...1E..71B}
{Bellm}, E., \& {Kulkarni}, S. 2017, Nature Astronomy, 1, 0071,
  \dodoi{10.1038/s41550-017-0071}

\bibitem[{{Bellm} {et~al.}(2019){Bellm}, {Kulkarni}, {Graham}, {Dekany},
  {Smith}, {Riddle}, {Masci}, {Helou}, {Prince}, {Adams}, {Barbarino},
  {Barlow}, {Bauer}, {Beck}, {Belicki}, {Biswas}, {Blagorodnova}, {Bodewits},
  {Bolin}, {Brinnel}, {Brooke}, {Bue}, {Bulla}, {Burruss}, {Cenko}, {Chang},
  {Connolly}, {Coughlin}, {Cromer}, {Cunningham}, {De}, {Delacroix}, {Desai},
  {Duev}, {Eadie}, {Farnham}, {Feeney}, {Feindt}, {Flynn}, {Franckowiak},
  {Frederick}, {Fremling}, {Gal-Yam}, {Gezari}, {Giomi}, {Goldstein},
  {Golkhou}, {Goobar}, {Groom}, {Hacopians}, {Hale}, {Henning}, {Ho}, {Hover},
  {Howell}, {Hung}, {Huppenkothen}, {Imel}, {Ip}, {Ivezi{\'c}}, {Jackson},
  {Jones}, {Juric}, {Kasliwal}, {Kaspi}, {Kaye}, {Kelley}, {Kowalski},
  {Kramer}, {Kupfer}, {Landry}, {Laher}, {Lee}, {Lin}, {Lin}, {Lunnan},
  {Giomi}, {Mahabal}, {Mao}, {Miller}, {Monkewitz}, {Murphy}, {Ngeow},
  {Nordin}, {Nugent}, {Ofek}, {Patterson}, {Penprase}, {Porter}, {Rauch},
  {Rebbapragada}, {Reiley}, {Rigault}, {Rodriguez}, {van Roestel}, {Rusholme},
  {van Santen}, {Schulze}, {Shupe}, {Singer}, {Soumagnac}, {Stein}, {Surace},
  {Sollerman}, {Szkody}, {Taddia}, {Terek}, {Van Sistine}, {van Velzen},
  {Vestrand}, {Walters}, {Ward}, {Ye}, {Yu}, {Yan}, \&
  {Zolkower}}]{2019PASP..131a8002B}
{Bellm}, E.~C., {Kulkarni}, S.~R., {Graham}, M.~J., {et~al.} 2019, \pasp, 131,
  018002, \dodoi{10.1088/1538-3873/aaecbe}

\bibitem[{{Bulla}(2019)}]{2019MNRAS.489.5037B}
{Bulla}, M. 2019, \mnras, 489, 5037, \dodoi{10.1093/mnras/stz2495}

\bibitem[{{Carbone} \& {Corsi}(2018)}]{2018ApJ...867..135C}
{Carbone}, D., \& {Corsi}, A. 2018, \apj, 867, 135,
  \dodoi{10.3847/1538-4357/aae583}

\bibitem[{{Carbone} \& {Corsi}(2020)}]{2020ApJ...889...36C}
---. 2020, \apj, 889, 36, \dodoi{10.3847/1538-4357/ab6227}

\bibitem[{{Cardelli} {et~al.}(1989){Cardelli}, {Clayton}, \&
  {Mathis}}]{1989ApJ...345..245C}
{Cardelli}, J.~A., {Clayton}, G.~C., \& {Mathis}, J.~S. 1989, \apj, 345, 245,
  \dodoi{10.1086/167900}

\bibitem[{{Chollet}(2016)}]{2016arXiv161002357C}
{Chollet}, F. 2016, arXiv e-prints, arXiv:1610.02357,
  \dodoi{10.48550/arXiv.1610.02357}

\bibitem[{Chollet {et~al.}(2015)}]{chollet2015keras}
Chollet, F., {et~al.} 2015, Keras, \url{https://keras.io}

\bibitem[{{Coughlin} {et~al.}(2023){Coughlin}, {Bloom}, {Nir}, {Antier}, {du
  Laz}, {van der Walt}, {Crellin-Quick}, {Culino}, {Duev}, {Goldstein},
  {Healy}, {Karambelkar}, {Lilleboe}, {Shin}, {Singer}, {Ahumada}, {Anand},
  {Bellm}, {Dekany}, {Graham}, {Kasliwal}, {Kostadinova}, {Kiendrebeogo},
  {Kulkarni}, {Jenkins}, {LeBaron}, {Mahabal}, {Neill}, {Parazin}, {Peloton},
  {Perley}, {Riddle}, {Rusholme}, {van Santen}, {Sollerman}, {Stein}, {Turpin},
  {Wold}, {Amat}, {Bonnefon}, {Bonnefoy}, {Flament}, {Kerkow}, {Kishore},
  {Jani}, {Mahanty}, {Liu}, {Llinares}, {Makarison}, {Olli{\'e}ric}, {Perez},
  {Pont}, \& {Sharma}}]{2023ApJS..267...31C}
{Coughlin}, M.~W., {Bloom}, J.~S., {Nir}, G., {et~al.} 2023, \apjs, 267, 31,
  \dodoi{10.3847/1538-4365/acdee1}

\bibitem[{{Coulter} {et~al.}(2017){Coulter}, {Foley}, {Kilpatrick}, {Drout},
  {Piro}, {Shappee}, {Siebert}, {Simon}, {Ulloa}, {Kasen}, {Madore},
  {Murguia-Berthier}, {Pan}, {Prochaska}, {Ramirez-Ruiz}, {Rest}, \&
  {Rojas-Bravo}}]{2017Sci...358.1556C}
{Coulter}, D.~A., {Foley}, R.~J., {Kilpatrick}, C.~D., {et~al.} 2017, Science,
  358, 1556, \dodoi{10.1126/science.aap9811}

\bibitem[{{Cranmer} {et~al.}(2021){Cranmer}, {Melchior}, \&
  {Nord}}]{2021arXiv210609761C}
{Cranmer}, M., {Melchior}, P., \& {Nord}, B. 2021, arXiv e-prints,
  arXiv:2106.09761.
\newblock \doarXiv{2106.09761}

\bibitem[{{Dekany} {et~al.}(2020){Dekany}, {Smith}, {Riddle}, {Feeney},
  {Porter}, {Hale}, {Zolkower}, {Belicki}, {Kaye}, {Henning}, {Walters},
  {Cromer}, {Delacroix}, {Rodriguez}, {Reiley}, {Mao}, {Hover}, {Murphy},
  {Burruss}, {Baker}, {Kowalski}, {Reif}, {Mueller}, {Bellm}, {Graham}, \&
  {Kulkarni}}]{2020PASP..132c8001D}
{Dekany}, R., {Smith}, R.~M., {Riddle}, R., {et~al.} 2020, \pasp, 132, 038001,
  \dodoi{10.1088/1538-3873/ab4ca2}

\bibitem[{{Drout} {et~al.}(2017){Drout}, {Piro}, {Shappee}, {Kilpatrick},
  {Simon}, {Contreras}, {Coulter}, {Foley}, {Siebert}, {Morrell}, {Boutsia},
  {Di Mille}, {Holoien}, {Kasen}, {Kollmeier}, {Madore}, {Monson},
  {Murguia-Berthier}, {Pan}, {Prochaska}, {Ramirez-Ruiz}, {Rest}, {Adams},
  {Alatalo}, {Ba{\~n}ados}, {Baughman}, {Beers}, {Bernstein}, {Bitsakis},
  {Campillay}, {Hansen}, {Higgs}, {Ji}, {Maravelias}, {Marshall}, {Moni Bidin},
  {Prieto}, {Rasmussen}, {Rojas-Bravo}, {Strom}, {Ulloa},
  {Vargas-Gonz{\'a}lez}, {Wan}, \& {Whitten}}]{2017Sci...358.1570D}
{Drout}, M.~R., {Piro}, A.~L., {Shappee}, B.~J., {et~al.} 2017, Science, 358,
  1570, \dodoi{10.1126/science.aaq0049}

\bibitem[{{F{\"o}rster} {et~al.}(2021){F{\"o}rster}, {Cabrera-Vives},
  {Castillo-Navarrete}, {Est{\'e}vez}, {S{\'a}nchez-S{\'a}ez}, {Arredondo},
  {Bauer}, {Carrasco-Davis}, {Catelan}, {Elorrieta}, {Eyheramendy}, {Huijse},
  {Pignata}, {Reyes}, {Reyes}, {Rodr{\'\i}guez-Mancini}, {Ruz-Mieres},
  {Valenzuela}, {{\'A}lvarez-Maldonado}, {Astorga}, {Borissova}, {Clocchiatti},
  {De Cicco}, {Donoso-Oliva}, {Hern{\'a}ndez-Garc{\'\i}a}, {Graham},
  {Jord{\'a}n}, {Kurtev}, {Mahabal}, {Maureira}, {Mu{\~n}oz-Arancibia},
  {Molina-Ferreiro}, {Moya}, {Palma}, {P{\'e}rez-Carrasco}, {Protopapas},
  {Romero}, {Sabatini-Gacitua}, {S{\'a}nchez}, {San Mart{\'\i}n},
  {Sep{\'u}lveda-Cobo}, {Vera}, \& {Vergara}}]{2021AJ....161..242F}
{F{\"o}rster}, F., {Cabrera-Vives}, G., {Castillo-Navarrete}, E., {et~al.}
  2021, \aj, 161, 242, \dodoi{10.3847/1538-3881/abe9bc}

\bibitem[{{Graham} {et~al.}(2019){Graham}, {Kulkarni}, {Bellm}, {Adams},
  {Barbarino}, {Blagorodnova}, {Bodewits}, {Bolin}, {Brady}, {Cenko}, {Chang},
  {Coughlin}, {De}, {Eadie}, {Farnham}, {Feindt}, {Franckowiak}, {Fremling},
  {Gezari}, {Ghosh}, {Goldstein}, {Golkhou}, {Goobar}, {Ho}, {Huppenkothen},
  {Ivezi{\'c}}, {Jones}, {Juric}, {Kaplan}, {Kasliwal}, {Kelley}, {Kupfer},
  {Lee}, {Lin}, {Lunnan}, {Mahabal}, {Miller}, {Ngeow}, {Nugent}, {Ofek},
  {Prince}, {Rauch}, {van Roestel}, {Schulze}, {Singer}, {Sollerman}, {Taddia},
  {Yan}, {Ye}, {Yu}, {Barlow}, {Bauer}, {Beck}, {Belicki}, {Biswas}, {Brinnel},
  {Brooke}, {Bue}, {Bulla}, {Burruss}, {Connolly}, {Cromer}, {Cunningham},
  {Dekany}, {Delacroix}, {Desai}, {Duev}, {Feeney}, {Flynn}, {Frederick},
  {Gal-Yam}, {Giomi}, {Groom}, {Hacopians}, {Hale}, {Helou}, {Henning},
  {Hover}, {Hillenbrand}, {Howell}, {Hung}, {Imel}, {Ip}, {Jackson}, {Kaspi},
  {Kaye}, {Kowalski}, {Kramer}, {Kuhn}, {Landry}, {Laher}, {Mao}, {Masci},
  {Monkewitz}, {Murphy}, {Nordin}, {Patterson}, {Penprase}, {Porter},
  {Rebbapragada}, {Reiley}, {Riddle}, {Rigault}, {Rodriguez}, {Rusholme}, {van
  Santen}, {Shupe}, {Smith}, {Soumagnac}, {Stein}, {Surace}, {Szkody}, {Terek},
  {Van Sistine}, {van Velzen}, {Vestrand}, {Walters}, {Ward}, {Zhang}, \&
  {Zolkower}}]{2019PASP..131g8001G}
{Graham}, M.~J., {Kulkarni}, S.~R., {Bellm}, E.~C., {et~al.} 2019, \pasp, 131,
  078001, \dodoi{10.1088/1538-3873/ab006c}

\bibitem[{{Guy} {et~al.}(2007){Guy}, {Astier}, {Baumont}, {Hardin}, {Pain},
  {Regnault}, {Basa}, {Carlberg}, {Conley}, {Fabbro}, {Fouchez}, {Hook},
  {Howell}, {Perrett}, {Pritchet}, {Rich}, {Sullivan}, {Antilogus}, {Aubourg},
  {Bazin}, {Bronder}, {Filiol}, {Palanque-Delabrouille}, {Ripoche}, \&
  {Ruhlmann-Kleider}}]{2007A&A...466...11G}
{Guy}, J., {Astier}, P., {Baumont}, S., {et~al.} 2007, \aap, 466, 11,
  \dodoi{10.1051/0004-6361:20066930}

\bibitem[{{Ho} {et~al.}(2020){Ho}, {Perley}, {Kulkarni}, {Dong}, {De},
  {Chandra}, {Andreoni}, {Bellm}, {Burdge}, {Coughlin}, {Dekany}, {Feeney},
  {Frederiks}, {Fremling}, {Golkhou}, {Graham}, {Hale}, {Helou}, {Horesh},
  {Kasliwal}, {Laher}, {Masci}, {Miller}, {Porter}, {Ridnaia}, {Rusholme},
  {Shupe}, {Soumagnac}, \& {Svinkin}}]{2020ApJ...895...49H}
{Ho}, A. Y.~Q., {Perley}, D.~A., {Kulkarni}, S.~R., {et~al.} 2020, \apj, 895,
  49, \dodoi{10.3847/1538-4357/ab8bcf}

\bibitem[{{Ho} {et~al.}(2022){Ho}, {Margalit}, {Bremer}, {Perley}, {Yao},
  {Dobie}, {Kaplan}, {O'Brien}, {Petitpas}, \& {Zic}}]{2022ApJ...932..116H}
{Ho}, A. Y.~Q., {Margalit}, B., {Bremer}, M., {et~al.} 2022, \apj, 932, 116,
  \dodoi{10.3847/1538-4357/ac4e97}

\bibitem[{{Ho} {et~al.}(2023){Ho}, {Perley}, {Gal-Yam}, {Lunnan}, {Sollerman},
  {Schulze}, {Das}, {Dobie}, {Yao}, {Fremling}, {Adams}, {Anand}, {Andreoni},
  {Bellm}, {Bruch}, {Burdge}, {Castro-Tirado}, {Dahiwale}, {De}, {Dekany},
  {Drake}, {Duev}, {Graham}, {Helou}, {Kaplan}, {Karambelkar}, {Kasliwal},
  {Kool}, {Kulkarni}, {Mahabal}, {Medford}, {Miller}, {Nordin}, {Ofek},
  {Petitpas}, {Riddle}, {Sharma}, {Smith}, {Stewart}, {Taggart}, {Tartaglia},
  {Tzanidakis}, \& {Winters}}]{2023ApJ...949..120H}
{Ho}, A. Y.~Q., {Perley}, D.~A., {Gal-Yam}, A., {et~al.} 2023, \apj, 949, 120,
  \dodoi{10.3847/1538-4357/acc533}

\bibitem[{{Ishida} {et~al.}(2019){Ishida}, {Beck}, {Gonz{\'a}lez-Gait{\'a}n},
  {de Souza}, {Krone-Martins}, {Barrett}, {Kennamer}, {Vilalta}, {Burgess},
  {Quint}, {Vitorelli}, {Mahabal}, \& {Gangler}}]{2019MNRAS.483....2I}
{Ishida}, E.~E.~O., {Beck}, R., {Gonz{\'a}lez-Gait{\'a}n}, S., {et~al.} 2019,
  \mnras, 483, 2, \dodoi{10.1093/mnras/sty3015}

\bibitem[{{Kasen} {et~al.}(2017){Kasen}, {Metzger}, {Barnes}, {Quataert}, \&
  {Ramirez-Ruiz}}]{2017Natur.551...80K}
{Kasen}, D., {Metzger}, B., {Barnes}, J., {Quataert}, E., \& {Ramirez-Ruiz}, E.
  2017, \nat, 551, 80, \dodoi{10.1038/nature24453}

\bibitem[{{Kasliwal} {et~al.}(2017){Kasliwal}, {Nakar}, {Singer}, {Kaplan},
  {Cook}, {Van Sistine}, {Lau}, {Fremling}, {Gottlieb}, {Jencson}, {Adams},
  {Feindt}, {Hotokezaka}, {Ghosh}, {Perley}, {Yu}, {Piran}, {Allison},
  {Anupama}, {Balasubramanian}, {Bannister}, {Bally}, {Barnes}, {Barway},
  {Bellm}, {Bhalerao}, {Bhattacharya}, {Blagorodnova}, {Bloom}, {Brady},
  {Cannella}, {Chatterjee}, {Cenko}, {Cobb}, {Copperwheat}, {Corsi}, {De},
  {Dobie}, {Emery}, {Evans}, {Fox}, {Frail}, {Frohmaier}, {Goobar}, {Hallinan},
  {Harrison}, {Helou}, {Hinderer}, {Ho}, {Horesh}, {Ip}, {Itoh}, {Kasen},
  {Kim}, {Kuin}, {Kupfer}, {Lynch}, {Madsen}, {Mazzali}, {Miller}, {Mooley},
  {Murphy}, {Ngeow}, {Nichols}, {Nissanke}, {Nugent}, {Ofek}, {Qi}, {Quimby},
  {Rosswog}, {Rusu}, {Sadler}, {Schmidt}, {Sollerman}, {Steele}, {Williamson},
  {Xu}, {Yan}, {Yatsu}, {Zhang}, \& {Zhao}}]{2017Sci...358.1559K}
{Kasliwal}, M.~M., {Nakar}, E., {Singer}, L.~P., {et~al.} 2017, Science, 358,
  1559, \dodoi{10.1126/science.aap9455}

\bibitem[{{Kasliwal} {et~al.}(2020){Kasliwal}, {Anand}, {Ahumada}, {Stein},
  {Carracedo}, {Andreoni}, {Coughlin}, {Singer}, {Kool}, {De}, {Kumar},
  {AlMualla}, {Yao}, {Bulla}, {Dobie}, {Reusch}, {Perley}, {Cenko}, {Bhalerao},
  {Kaplan}, {Sollerman}, {Goobar}, {Copperwheat}, {Bellm}, {Anupama}, {Corsi},
  {Nissanke}, {Agudo}, {Bagdasaryan}, {Barway}, {Belicki}, {Bloom}, {Bolin},
  {Buckley}, {Burdge}, {Burruss}, {Caballero-Garc{\'\i}a}, {Cannella},
  {Castro-Tirado}, {Cook}, {Cooke}, {Cunningham}, {Dahiwale}, {Deshmukh},
  {Dichiara}, {Duev}, {Dutta}, {Feeney}, {Franckowiak}, {Frederick},
  {Fremling}, {Gal-Yam}, {Gatkine}, {Ghosh}, {Goldstein}, {Golkhou}, {Graham},
  {Graham}, {Hankins}, {Helou}, {Hu}, {Ip}, {Jaodand}, {Karambelkar}, {Kong},
  {Kowalski}, {Khandagale}, {Kulkarni}, {Kumar}, {Laher}, {Li}, {Mahabal},
  {Masci}, {Miller}, {Mogotsi}, {Mohite}, {Mooley}, {Mroz}, {Newman}, {Ngeow},
  {Oates}, {Patil}, {Pandey}, {Pavana}, {Pian}, {Riddle},
  {S{\'a}nchez-Ram{\'\i}rez}, {Sharma}, {Singh}, {Smith}, {Soumagnac},
  {Taggart}, {Tan}, {Tzanidakis}, {Troja}, {Valeev}, {Walters}, {Waratkar},
  {Webb}, {Yu}, {Zhang}, {Zhou}, \& {Zolkower}}]{2020ApJ...905..145K}
{Kasliwal}, M.~M., {Anand}, S., {Ahumada}, T., {et~al.} 2020, \apj, 905, 145,
  \dodoi{10.3847/1538-4357/abc335}

\bibitem[{{Kennamer} {et~al.}(2020){Kennamer}, {Ishida}, {Gonzalez-Gaitan}, {de
  Souza}, {Ihler}, {Ponder}, {Vilalta}, {Moller}, {Jones}, {Dai},
  {Krone-Martins}, {Quint}, {Sreejith}, {Malz}, \&
  {Galbany}}]{2020arXiv201005941K}
{Kennamer}, N., {Ishida}, E. E.~O., {Gonzalez-Gaitan}, S., {et~al.} 2020, arXiv
  e-prints, arXiv:2010.05941.
\newblock \doarXiv{2010.05941}

\bibitem[{{Kiendrebeogo} {et~al.}(2023){Kiendrebeogo}, {Farah}, {Foley},
  {Gray}, {Kunert}, {Puecher}, {Toivonen}, {VandenBerg}, {Anand}, {Ahumada},
  {Karambelkar}, {Coughlin}, {Dietrich}, {Kam}, {Pang}, {Singer}, \&
  {Sravan}}]{2023ApJ...958..158K}
{Kiendrebeogo}, R.~W., {Farah}, A.~M., {Foley}, E.~M., {et~al.} 2023, \apj,
  958, 158, \dodoi{10.3847/1538-4357/acfcb1}

\bibitem[{{Kochanek} {et~al.}(2017){Kochanek}, {Shappee}, {Stanek}, {Holoien},
  {Thompson}, {Prieto}, {Dong}, {Shields}, {Will}, {Britt}, {Perzanowski}, \&
  {Pojma{\'n}ski}}]{2017PASP..129j4502K}
{Kochanek}, C.~S., {Shappee}, B.~J., {Stanek}, K.~Z., {et~al.} 2017, \pasp,
  129, 104502, \dodoi{10.1088/1538-3873/aa80d9}

\bibitem[{{Lattimer} \& {Schramm}(1974)}]{1974ApJ...192L.145L}
{Lattimer}, J.~M., \& {Schramm}, D.~N. 1974, \apjl, 192, L145,
  \dodoi{10.1086/181612}

\bibitem[{{Li} \& {Paczy{\'n}ski}(1998)}]{1998ApJ...507L..59L}
{Li}, L.-X., \& {Paczy{\'n}ski}, B. 1998, \apjl, 507, L59,
  \dodoi{10.1086/311680}

\bibitem[{LSC {et~al.}(2015)LSC, Aasi, Abbott, Abbott, Abbott, Abernathy,
  Ackley, Adams, Adams, Addesso, Adhikari, Adya, Affeldt, Aggarwal, Aguiar,
  Ain, Ajith, Alemic, Allen, Amariutei, Anderson, Anderson, Arai, Araya,
  Arceneaux, Areeda, Ashton, Ast, Aston, Aufmuth, Aulbert, Aylott, Babak,
  Baker, Ballmer, Barayoga, Barbet, Barclay, Barish, Barker, Barr, Barsotti,
  Bartlett, Barton, Bartos, Bassiri, Batch, Baune, Behnke, Bell, Bell,
  Benacquista, Bergman, Bergmann, Berry, Betzwieser, Bhagwat, Bhandare,
  Bilenko, Billingsley, Birch, Biscans, Biwer, Blackburn, Blackburn, Blair,
  Blair, Bock, Bodiya, Bojtos, Bond, Bork, Born, Bose, Brady, Braginsky, Brau,
  Bridges, Brinkmann, Brooks, Brown, Brown, Brown, Buchman, Buikema, Buonanno,
  Cadonati, Bustillo, Camp, Cannon, Cao, Capano, Caride, Caudill, Cavagli{\`a},
  Cepeda, Chakraborty, Chalermsongsak, Chamberlin, Chao, Charlton, Chen, Cho,
  Cho, Chow, Christensen, Chu, Chung, Ciani, Clara, Clark, Collette, Cominsky,
  Constancio, Cook, Corbitt, Cornish, Corsi, Costa, Coughlin, Countryman,
  Couvares, Coward, Cowart, Coyne, Coyne, Craig, Creighton, Creighton, Cripe,
  Crowder, Cumming, Cunningham, Cutler, Dahl, Canton, Damjanic, Danilishin,
  Danzmann, Dartez, Dave, Daveloza, Davies, Daw, DeBra, Pozzo, Denker, Dent,
  Dergachev, DeRosa, DeSalvo, Dhurandhar, D´ıaz, Palma, Dojcinoski,
  Dominguez, Donovan, Dooley, Doravari, Douglas, Downes, Driggers, Du, Dwyer,
  Eberle, Edo, Edwards, Edwards, Effler, Eggenstein, Ehrens, Eichholz,
  Eikenberry, Essick, Etzel, Evans, Evans, Factourovich, Fairhurst, Fan, Fang,
  Farr, Farr, Favata, Fays, Fehrmann, Fejer, Feldbaum, Ferreira, Fisher, Frei,
  Freise, Frey, Fricke, Fritschel, Frolov, Fuentes-Tapia, Fulda, Fyffe, Gair,
  Gaonkar, Gehrels, Gergely´, Giaime, Giardina, Gleason, Goetz, Goetz, Gondan,
  Gonz{\'a}lez, Gordon, Gorodetsky, Gossan, Go{\ss}ler, Gr{\"a}f, Graff, Grant,
  Gras, Gray, Greenhalgh, Gretarsson, Grote, Grunewald, Guido, Guo, Gushwa,
  Gustafson, Gustafson, Hacker, Hall, Hammond, Hanke, Hanks, Hanna, Hannam,
  Hanson, Hardwick, Harry, Harry, Hart, Hartman, Haster, Haughian, Hee,
  Heintze, Heinzel, Hendry, Heng, Heptonstall, Heurs, Hewitson, Hild, Hoak,
  Hodge, Hollitt, Holt, Hopkins, Hosken, Hough, Houston, Howell, Hu, Huerta,
  Hughey, Husa, Huttner, Huynh, Huynh-Dinh, Idrisy, Indik, Ingram, Inta, Islas,
  Isler, Isogai, Iyer, Izumi, Jacobson, Jang, Jawahar, Ji, Jim{\'e}nez-Forteza,
  Johnson, Jones, Jones, Ju, Haris, Kalogera, Kandhasamy, Kang, Kanner,
  Katsavounidis, Katzman, Kaufer, Kaufer, Kaur, Kawabe, Kawazoe, Keiser,
  Keitel, Kelley, Kells, Keppel, Key, Khalaidovski, Khalili, Khazanov, Kim,
  Kim, Kim, Kim, Kim, King, King, Kinzel, Kissel, Klimenko, Kline, Koehlenbeck,
  Kokeyama, Kondrashov, Korobko, Korth, Kozak, Kringel, Krishnan, Krueger,
  Kuehn, Kumar, Kumar, Kuo, Landry, Lantz, Larson, Lasky, Lazzarini, Lazzaro,
  Le, Leaci, Leavey, Lebigot, Lee, Lee, Lee, Leong, Levin, Levine, Lewis, Li,
  Libbrecht, Libson, Lin, Littenberg, Lockerbie, Lockett, Logue, Lombardi,
  Lormand, Lough, Lubinski, L{\"u}ck, Lundgren, Lynch, Ma, Macarthur,
  MacDonald, Machenschalk, MacInnis, Macleod, Maga{\~n}a-Sandoval, Magee,
  Mageswaran, Maglione, Mailand, Mandel, Mandic, Mangano, Mansell, M{\'a}rka,
  M{\'a}rka, Markosyan, Maros, Martin, Martin, Martynov, Marx, Mason,
  Massinger, Matichard, Matone, Mavalvala, Mazumder, Mazzolo, McCarthy,
  McClelland, McCormick, McGuire, McIntyre, McIver, McLin, McWilliams, Meadors,
  Meinders, Melatos, Mendell, Mercer, Meshkov, Messenger, Meyers, Miao,
  Middleton, Mikhailov, Miller, Miller, Millhouse, Ming, Mirshekari, Mishra,
  Mitra, Mitrofanov, Mitselmakher, Mittleman, Moe, Mohanty, Mohapatra, Moore,
  Moraru, Moreno, Morriss, Mossavi, Mow-Lowry, Mueller, Mueller, Mukherjee,
  Mullavey, Munch, Murphy, Murray, Mytidis, Nash, Nayak, Necula, Nedkova,
  Newton, Nguyen, Nielsen, Nissanke, Nitz, Nolting, Normandin, Nuttall,
  Ochsner, O'Dell, Oelker, Ogin, Oh, Oh, Ohme, Oppermann, Oram, O'Reilly,
  Ortega, O'Shaughnessy, Osthelder, Ott, Ottaway, Ottens, Overmier, Owen,
  Padilla, Pai, Pai, Palashov, Pal-Singh, Pan, Pankow, Pannarale, Pant, Papa,
  Paris, Patrick, Pedraza, Pekowsky, Pele, Penn, Perreca, Phelps, Pierro,
  Pinto, Pitkin, Poeld, Post, Poteomkin, Powell, Prasad, Predoi, Premachandra,
  Prestegard, Price, Principe, Privitera, Prix, Prokhorov, Puncken, P{\"u}rrer,
  Qin, Quetschke, Quintero, Quiroga, Quitzow-James, Raab, Rabeling, Radkins,
  Raffai, Raja, Rajalakshmi, Rakhmanov, Ramirez, Raymond, Reed, Reid, Reitze,
  Reula, Riles, Robertson, Robie, Rollins, Roma, Romano, Romanov, Romie, Rowan,
  R{\"u}diger, Ryan, Sachdev, Sadecki, Sadeghian, Saleem, Salemi, Sammut,
  Sandberg, Sanders, Sannibale, Santiago-Prieto, Sathyaprakash, Saulson,
  Savage, Sawadsky, Scheuer, Schilling, Schmidt, Schnabel, Schofield,
  Schreiber, Schuette, Schutz, Scott, Scott, Sellers, Sengupta, Sergeev, Serna,
  Sevigny, Shaddock, Shahriar, Shaltev, Shao, Shapiro, Shawhan, Shoemaker,
  Sidery, Siemens, Sigg, Silva, Simakov, Singer, Singer, Singh, Sintes,
  Slagmolen, Smith, Smith, Smith, Smith-Lefebvre, Son, Sorazu, Souradeep,
  Staley, Stebbins, Steinke, Steinlechner, Steinlechner, Steinmeyer, Stephens,
  Steplewski, Stevenson, Stone, Strain, Strigin, Sturani, Stuver, Summerscales,
  Sutton, Szczepanczyk, Szeifert, Talukder, Tanner, T{\'a}pai, Tarabrin,
  Taracchini, Taylor, Tellez, Theeg, Thirugnanasambandam, Thomas, Thomas,
  Thorne, Thorne, Thrane, Tiwari, Tomlinson, Torres, Torrie, Traylor, Tse,
  Tshilumba, Ugolini, Unnikrishnan, Urban, Usman, Vahlbruch, Vajente, Valdes,
  Vallisneri, van Veggel, Vass, Vaulin, Vecchio, Veitch, Veitch, Venkateswara,
  Vincent-Finley, Vitale, Vo, Vorvick, Vousden, Vyatchanin, Wade, Wade, Wade,
  Walker, Wallace, Walsh, Wang, Wang, Wang, Ward, Warner, Was, Weaver, Weinert,
  Weinstein, Weiss, Welborn, Wen, Wessels, Westphal, Wette, Whelan, Whitcomb,
  White, Whiting, Wilkinson, Williams, Williams, Williamson, Willis, Willke,
  Wimmer, Winkler, Wipf, Wittel, Woan, Worden, Xie, Yablon, Yakushin, Yam,
  Yamamoto, Yancey, Yang, Zanolin, Zhang, Zhang, Zhang, Zhang, Zhao, Zhou, Zhu,
  Zucker, Zuraw, \& Zweizig}]{LIGO_detector}
LSC, Aasi, J., Abbott, B.~P., {et~al.} 2015, Classical and Quantum Gravity, 32,
  074001, \dodoi{10.1088/0264-9381/32/7/074001}

\bibitem[{{Mandhai} {et~al.}(2018){Mandhai}, {Tanvir}, {Lamb}, {Levan}, \&
  {Tsang}}]{2018Galax...6..130M}
{Mandhai}, S., {Tanvir}, N., {Lamb}, G., {Levan}, A., \& {Tsang}, D. 2018,
  Galaxies, 6, 130, \dodoi{10.3390/galaxies6040130}

\bibitem[{{Masci} {et~al.}(2019){Masci}, {Laher}, {Rusholme}, {Shupe}, {Groom},
  {Surace}, {Jackson}, {Monkewitz}, {Beck}, {Flynn}, {Terek}, {Landry},
  {Hacopians}, {Desai}, {Howell}, {Brooke}, {Imel}, {Wachter}, {Ye}, {Lin},
  {Cenko}, {Cunningham}, {Rebbapragada}, {Bue}, {Miller}, {Mahabal}, {Bellm},
  {Patterson}, {Juri{\'c}}, {Golkhou}, {Ofek}, {Walters}, {Graham}, {Kasliwal},
  {Dekany}, {Kupfer}, {Burdge}, {Cannella}, {Barlow}, {Van Sistine}, {Giomi},
  {Fremling}, {Blagorodnova}, {Levitan}, {Riddle}, {Smith}, {Helou}, {Prince},
  \& {Kulkarni}}]{2019PASP..131a8003M}
{Masci}, F.~J., {Laher}, R.~R., {Rusholme}, B., {et~al.} 2019, \pasp, 131,
  018003, \dodoi{10.1088/1538-3873/aae8ac}

\bibitem[{{Metzger}(2019)}]{2019LRR....23....1M}
{Metzger}, B.~D. 2019, Living Reviews in Relativity, 23, 1,
  \dodoi{10.1007/s41114-019-0024-0}

\bibitem[{{Metzger} {et~al.}(2010){Metzger}, {Mart{\'\i}nez-Pinedo}, {Darbha},
  {Quataert}, {Arcones}, {Kasen}, {Thomas}, {Nugent}, {Panov}, \&
  {Zinner}}]{2010MNRAS.406.2650M}
{Metzger}, B.~D., {Mart{\'\i}nez-Pinedo}, G., {Darbha}, S., {et~al.} 2010,
  \mnras, 406, 2650, \dodoi{10.1111/j.1365-2966.2010.16864.x}

\bibitem[{Mnih {et~al.}(2013)Mnih, Kavukcuoglu, Silver, Graves, Antonoglou,
  Wierstra, \& Riedmiller}]{Mnih2013PlayingAW}
Mnih, V., Kavukcuoglu, K., Silver, D., {et~al.} 2013, ArXiv, abs/1312.5602

\bibitem[{{Mnih} {et~al.}(2016){Mnih}, {Puigdom{\`e}nech Badia}, {Mirza},
  {Graves}, {Lillicrap}, {Harley}, {Silver}, \&
  {Kavukcuoglu}}]{2016arXiv160201783M}
{Mnih}, V., {Puigdom{\`e}nech Badia}, A., {Mirza}, M., {et~al.} 2016, arXiv
  e-prints, arXiv:1602.01783.
\newblock \doarXiv{1602.01783}

\bibitem[{{M{\"o}ller} {et~al.}(2021){M{\"o}ller}, {Peloton}, {Ishida},
  {Arnault}, {Bachelet}, {Blaineau}, {Boutigny}, {Chauhan}, {Gangler},
  {Hernandez}, {Hrivnac}, {Leoni}, {Leroy}, {Moniez}, {Pateyron}, {Ramparison},
  {Turpin}, {Ansari}, {Allam}, {Bajat}, {Biswas}, {Boucaud}, {Bregeon},
  {Campagne}, {Cohen-Tanugi}, {Coleiro}, {Dornic}, {Fouchez}, {Godet}, {Gris},
  {Karpov}, {Nebot Gomez-Moran}, {Neveu}, {Plaszczynski}, {Savchenko}, \&
  {Webb}}]{2021MNRAS.501.3272M}
{M{\"o}ller}, A., {Peloton}, J., {Ishida}, E. E.~O., {et~al.} 2021, \mnras,
  501, 3272, \dodoi{10.1093/mnras/staa3602}

\bibitem[{{Narayan} {et~al.}(1992){Narayan}, {Paczynski}, \&
  {Piran}}]{1992ApJ...395L..83N}
{Narayan}, R., {Paczynski}, B., \& {Piran}, T. 1992, \apjl, 395, L83,
  \dodoi{10.1086/186493}

\bibitem[{{Nordin} {et~al.}(2019){Nordin}, {Brinnel}, {van Santen}, {Bulla},
  {Feindt}, {Franckowiak}, {Fremling}, {Gal-Yam}, {Giomi}, {Kowalski},
  {Mahabal}, {Miranda}, {Rauch}, {Reusch}, {Rigault}, {Schulze}, {Sollerman},
  {Stein}, {Yaron}, {van Velzen}, \& {Ward}}]{2019A&A...631A.147N}
{Nordin}, J., {Brinnel}, V., {van Santen}, J., {et~al.} 2019, \aap, 631, A147,
  \dodoi{10.1051/0004-6361/201935634}

\bibitem[{pandas~development team(2020)}]{reback2020pandas}
pandas~development team, T. 2020, pandas-dev/pandas: Pandas, latest,  Zenodo,
  \dodoi{10.5281/zenodo.3509134}

\bibitem[{{Pang} {et~al.}(2023){Pang}, {Dietrich}, {Coughlin}, {Bulla}, {Tews},
  {Almualla}, {Barna}, {Kiendrebeogo}, {Kunert}, {Mansingh}, {Reed}, {Sravan},
  {Toivonen}, {Antier}, {VandenBerg}, {Heinzel}, {Nedora}, {Salehi}, {Sharma},
  {Somasundaram}, \& {Van Den Broeck}}]{2023NatCo..14.8352P}
{Pang}, P. T.~H., {Dietrich}, T., {Coughlin}, M.~W., {et~al.} 2023, Nature
  Communications, 14, 8352, \dodoi{10.1038/s41467-023-43932-6}

\bibitem[{Pedregosa {et~al.}(2011)Pedregosa, Varoquaux, Gramfort, Michel,
  Thirion, Grisel, Blondel, Prettenhofer, Weiss, Dubourg, Vanderplas, Passos,
  Cournapeau, Brucher, Perrot, \& Duchesnay}]{scikit-learn}
Pedregosa, F., Varoquaux, G., Gramfort, A., {et~al.} 2011, Journal of Machine
  Learning Research, 12, 2825

\bibitem[{{Petrov} {et~al.}(2022){Petrov}, {Singer}, {Coughlin}, {Kumar},
  {Almualla}, {Anand}, {Bulla}, {Dietrich}, {Foucart}, \&
  {Guessoum}}]{2022ApJ...924...54P}
{Petrov}, P., {Singer}, L.~P., {Coughlin}, M.~W., {et~al.} 2022, \apj, 924, 54,
  \dodoi{10.3847/1538-4357/ac366d}

\bibitem[{{Radice} {et~al.}(2018){Radice}, {Perego}, {Zappa}, \&
  {Bernuzzi}}]{2018ApJ...852L..29R}
{Radice}, D., {Perego}, A., {Zappa}, F., \& {Bernuzzi}, S. 2018, \apjl, 852,
  L29, \dodoi{10.3847/2041-8213/aaa402}

\bibitem[{Ross {et~al.}(2011)Ross, Gordon, \& Bagnell}]{pmlr-v15-ross11a}
Ross, S., Gordon, G., \& Bagnell, D. 2011, in Proceedings of Machine Learning
  Research, Vol.~15, Proceedings of the Fourteenth International Conference on
  Artificial Intelligence and Statistics, ed. G.~Gordon, D.~Dunson, \&
  M.~Dud{\'\i}k (Fort Lauderdale, FL, USA: PMLR), 627--635.
\newblock \url{https://proceedings.mlr.press/v15/ross11a.html}

\bibitem[{Rummery \& Niranjan(1994)}]{Rummery1994OnlineQU}
Rummery, G.~A., \& Niranjan, M. 1994

\bibitem[{{Ryan} {et~al.}(2020){Ryan}, {van Eerten}, {Piro}, \&
  {Troja}}]{2020ApJ...896..166R}
{Ryan}, G., {van Eerten}, H., {Piro}, L., \& {Troja}, E. 2020, \apj, 896, 166,
  \dodoi{10.3847/1538-4357/ab93cf}

\bibitem[{{Saha} {et~al.}(2014){Saha}, {Matheson}, {Snodgrass}, {Kececioglu},
  {Narayan}, {Seaman}, {Jenness}, \& {Axelrod}}]{2014SPIE.9149E..08S}
{Saha}, A., {Matheson}, T., {Snodgrass}, R., {et~al.} 2014, in Society of
  Photo-Optical Instrumentation Engineers (SPIE) Conference Series, Vol. 9149,
  Observatory Operations: Strategies, Processes, and Systems V, 914908,
  \dodoi{10.1117/12.2056988}

\bibitem[{{Schaul} {et~al.}(2015){Schaul}, {Quan}, {Antonoglou}, \&
  {Silver}}]{2015arXiv151105952S}
{Schaul}, T., {Quan}, J., {Antonoglou}, I., \& {Silver}, D. 2015, arXiv
  e-prints, arXiv:1511.05952.
\newblock \doarXiv{1511.05952}

\bibitem[{{Schlegel} {et~al.}(1998){Schlegel}, {Finkbeiner}, \&
  {Davis}}]{1998ApJ...500..525S}
{Schlegel}, D.~J., {Finkbeiner}, D.~P., \& {Davis}, M. 1998, \apj, 500, 525,
  \dodoi{10.1086/305772}

\bibitem[{{Singer} {et~al.}(2015){Singer}, {Kasliwal}, {Cenko}, {Perley},
  {Anderson}, {Anupama}, {Arcavi}, {Bhalerao}, {Bue}, {Cao}, {Connaughton},
  {Corsi}, {Cucchiara}, {Fender}, {Fox}, {Gehrels}, {Goldstein}, {Gorosabel},
  {Horesh}, {Hurley}, {Johansson}, {Kann}, {Kouveliotou}, {Huang}, {Kulkarni},
  {Masci}, {Nugent}, {Rau}, {Rebbapragada}, {Staley}, {Svinkin}, {Th{\"o}ne},
  {de Ugarte Postigo}, {Urata}, \& {Weinstein}}]{2015ApJ...806...52S}
{Singer}, L.~P., {Kasliwal}, M.~M., {Cenko}, S.~B., {et~al.} 2015, \apj, 806,
  52, \dodoi{10.1088/0004-637X/806/1/52}

\bibitem[{{Smartt} {et~al.}(2017){Smartt}, {Chen}, {Jerkstrand}, {Coughlin},
  {Kankare}, {Sim}, {Fraser}, {Inserra}, {Maguire}, {Chambers}, {Huber},
  {Kr{\"u}hler}, {Leloudas}, {Magee}, {Shingles}, {Smith}, {Young}, {Tonry},
  {Kotak}, {Gal-Yam}, {Lyman}, {Homan}, {Agliozzo}, {Anderson}, {Angus},
  {Ashall}, {Barbarino}, {Bauer}, {Berton}, {Botticella}, {Bulla}, {Bulger},
  {Cannizzaro}, {Cano}, {Cartier}, {Cikota}, {Clark}, {De Cia}, {Della Valle},
  {Denneau}, {Dennefeld}, {Dessart}, {Dimitriadis}, {Elias-Rosa}, {Firth},
  {Flewelling}, {Fl{\"o}rs}, {Franckowiak}, {Frohmaier}, {Galbany},
  {Gonz{\'a}lez-Gait{\'a}n}, {Greiner}, {Gromadzki}, {Guelbenzu},
  {Guti{\'e}rrez}, {Hamanowicz}, {Hanlon}, {Harmanen}, {Heintz}, {Heinze},
  {Hernandez}, {Hodgkin}, {Hook}, {Izzo}, {James}, {Jonker}, {Kerzendorf},
  {Klose}, {Kostrzewa-Rutkowska}, {Kowalski}, {Kromer}, {Kuncarayakti},
  {Lawrence}, {Lowe}, {Magnier}, {Manulis}, {Martin-Carrillo}, {Mattila},
  {McBrien}, {M{\"u}ller}, {Nordin}, {O'Neill}, {Onori}, {Palmerio},
  {Pastorello}, {Patat}, {Pignata}, {Podsiadlowski}, {Pumo}, {Prentice}, {Rau},
  {Razza}, {Rest}, {Reynolds}, {Roy}, {Ruiter}, {Rybicki}, {Salmon}, {Schady},
  {Schultz}, {Schweyer}, {Seitenzahl}, {Smith}, {Sollerman}, {Stalder},
  {Stubbs}, {Sullivan}, {Szegedi}, {Taddia}, {Taubenberger}, {Terreran}, {van
  Soelen}, {Vos}, {Wainscoat}, {Walton}, {Waters}, {Weiland}, {Willman},
  {Wiseman}, {Wright}, {Wyrzykowski}, \& {Yaron}}]{2017Natur.551...75S}
{Smartt}, S.~J., {Chen}, T.~W., {Jerkstrand}, A., {et~al.} 2017, \nat, 551, 75,
  \dodoi{10.1038/nature24303}

\bibitem[{Sravan(2024)}]{niharika_sravan_2024_10995342}
Sravan, N. 2024, {niharika-sravan/Pythia: Associated with ApJ article
  "Machine-directed gravitational-wave counterpart discovery"}, v0.0.1,
  Zenodo, \dodoi{10.5281/zenodo.10995342}

\bibitem[{{Sravan} {et~al.}(2021){Sravan}, {Graham}, {Fremling}, \&
  {Coughlin}}]{2021arXiv211205897S}
{Sravan}, N., {Graham}, M.~J., {Fremling}, C., \& {Coughlin}, M.~W. 2021, arXiv
  e-prints, arXiv:2112.05897.
\newblock \doarXiv{2112.05897}

\bibitem[{{Sravan} {et~al.}(2020){Sravan}, {Milisavljevic}, {Reynolds},
  {Lentner}, \& {Linvill}}]{2020ApJ...893..127S}
{Sravan}, N., {Milisavljevic}, D., {Reynolds}, J.~M., {Lentner}, G., \&
  {Linvill}, M. 2020, \apj, 893, 127, \dodoi{10.3847/1538-4357/ab8128}

\bibitem[{{Street} {et~al.}(2018){Street}, {Bowman}, {Saunders}, \&
  {Boroson}}]{2018SPIE10707E..11S}
{Street}, R.~A., {Bowman}, M., {Saunders}, E.~S., \& {Boroson}, T. 2018, in
  Society of Photo-Optical Instrumentation Engineers (SPIE) Conference Series,
  Vol. 10707, Software and Cyberinfrastructure for Astronomy V, 1070711,
  \dodoi{10.1117/12.2312293}

\bibitem[{Sutton \& Barto(2018)}]{sutton2018reinforcement}
Sutton, R.~S., \& Barto, A.~G. 2018, Reinforcement learning: An introduction
  (MIT press)

\bibitem[{{Symbalisty} \& {Schramm}(1982)}]{1982ApL....22..143S}
{Symbalisty}, E., \& {Schramm}, D.~N. 1982, \aplett, 22, 143

\bibitem[{Tesauro(1995)}]{10.1145/203330.203343}
Tesauro, G. 1995, Commun. ACM, 38, 58, \dodoi{10.1145/203330.203343}

\bibitem[{{Tonry} {et~al.}(2018){Tonry}, {Denneau}, {Heinze}, {Stalder},
  {Smith}, {Smartt}, {Stubbs}, {Weiland }, \& {Rest}}]{2018PASP..130f4505T}
{Tonry}, J.~L., {Denneau}, L., {Heinze}, A.~N., {et~al.} 2018, \pasp, 130,
  064505, \dodoi{10.1088/1538-3873/aabadf}

\bibitem[{Tsitsiklis \& Van~Roy(1997)}]{580874}
Tsitsiklis, J., \& Van~Roy, B. 1997, IEEE Transactions on Automatic Control,
  42, 674, \dodoi{10.1109/9.580874}

\bibitem[{{Wang} \& {Melchior}(2022)}]{2022MLS&T...3a5023W}
{Wang}, T., \& {Melchior}, P. 2022, Machine Learning: Science and Technology,
  3, 015023, \dodoi{10.1088/2632-2153/ac4d12}

\bibitem[{{Wang} {et~al.}(2016){Wang}, {Bapst}, {Heess}, {Mnih}, {Munos},
  {Kavukcuoglu}, \& {de Freitas}}]{2016arXiv161101224W}
{Wang}, Z., {Bapst}, V., {Heess}, N., {et~al.} 2016, arXiv e-prints,
  arXiv:1611.01224.
\newblock \doarXiv{1611.01224}

\bibitem[{Watkins(1989)}]{watkins1989learning}
Watkins, C. J. C.~H. 1989

\bibitem[{{W}es {M}c{K}inney(2010)}]{mckinney-proc-scipy-2010}
{W}es {M}c{K}inney. 2010, in {P}roceedings of the 9th {P}ython in {S}cience
  {C}onference, ed. {S}t\'efan van~der {W}alt \& {J}arrod {M}illman, 56 -- 61,
  \dodoi{10.25080/Majora-92bf1922-00a}

\bibitem[{{Williamson} {et~al.}(2019){Williamson}, {Modjaz}, \&
  {Bianco}}]{2019ApJ...880L..22W}
{Williamson}, M., {Modjaz}, M., \& {Bianco}, F.~B. 2019, \apjl, 880, L22,
  \dodoi{10.3847/2041-8213/ab2edb}

\end{thebibliography}
\bibliographystyle{aasjournal}



\end{document}